\documentclass[a4paper,bibliography=totoc,headings=standardclasses,chapterprefix=false,headsepline]{scrbook}
\pdfoutput=1
\usepackage[utf8]{inputenc} 
\usepackage[T1]{fontenc}
\usepackage{textcomp}
\usepackage{lmodern}
\usepackage[sc,osf]{mathpazo} 
\usepackage[UKenglish]{babel}

\usepackage{amsmath}
\usepackage{amssymb}
\usepackage{tensor}

\usepackage{relsize}

\usepackage{csquotes}
\usepackage{epigraph}
\renewcommand{\epsilon}{\varepsilon}

\usepackage{microtype}
\usepackage{tikz}

\usepackage[style=ext-numeric,date=year,url=false,articlein=false,backref=true]{biblatex}
\DeclareFieldFormat{pages}{#1} 
\DeclareFieldFormat[article]{volume}{\textbf{#1}} 
\DeclareFieldFormat[article]{journaltitle}{#1} 
\AtEveryBibitem{%
	\clearfield{number}
	\clearfield{issue}} 
\renewbibmacro*{pageref}{
	\iflistundef{pageref}{}
	{ $\hookrightarrow$ \ppspace\printlist[pageref][-\value{listtotal}]{pageref}}}

\usepackage{graphicx}
\usepackage{transparent}
\usepackage{titlepic}

\usepackage[colorlinks]{hyperref}

\title{Geometric post-Newtonian description of spin-half particles in curved spacetime }
\author{Ashkan Alibabaei}
\date{12th September 2022\\\textsmaller{(originally submitted 1st April 2022)}}
\titlepic{\includegraphics[width=\textwidth]{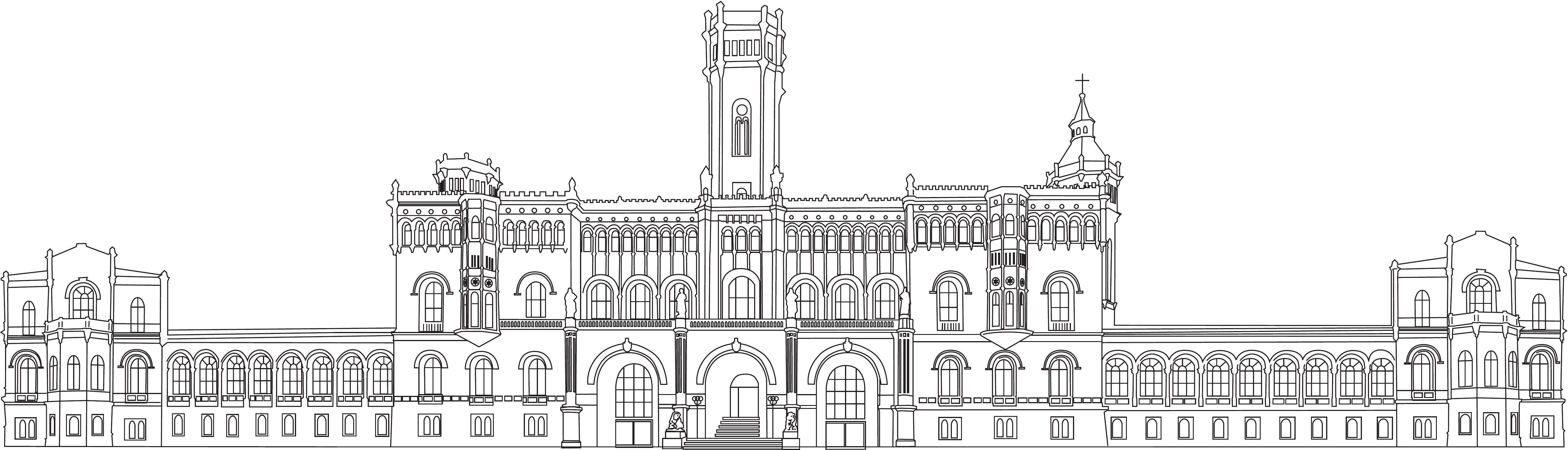}}

\begin{document}

\frontmatter

\begin{titlepage}
\makeatletter
    \begin{center}
        \Huge
        \scshape
\@title 

	\vspace{-.5 em}
{	\rule{.9\textwidth}{1.5pt}}
      \end{center}
      
\vfill

       \begin{center}
        { \huge Ashkan Alibabaei}
            
        \vfill
         
        {   
         \LARGE Master's Thesis }
         
         \vspace*{1ex}
         
         {\large (updated version)}
         \end{center}   
         
\vfill \vfill
  \begin{center}         
        \transparent{0.5}\includegraphics[width=\textwidth]{welfenschloss.pdf}
            \end{center}
            
          \vfill  
 \begin{center}         
        \Large
        Institute of Theoretical Physics\\
        Leibniz University Hannover, Germany
       
        \@date
            
    \end{center}
\makeatother    
\end{titlepage}

\thispagestyle{empty}
\noindent
\textbf{Examiners:}\\[.5em]
Prof.\ Dr.\ Domenico Giulini\\
Prof.\ Dr.\ Klemens Hammerer\\

\noindent
\textbf{Supervisors:}\\[.5em]
Prof.\ Dr.\ Domenico Giulini\\
Dr.\ Philip K.\ Schwartz\\

\vfill

\noindent
This is an updated version of the original version of my master's thesis, which has been adapted and corrected in various places.

\cleardoublepage

\chapter*{Abstract}

Einstein Equivalence Principle (EEP) requires all matter components to universally couple to gravity via a single common geometry: that of spacetime. This relates quantum theory with geometry as soon as interactions with gravity are considered. In this work, I study the geometric theory of coupling a spin-$\frac{1}{2}$ particle to gravity in a twofold expansion scheme: First with respect to the distance based on Fermi normal coordinates around a preferred worldline (e.g., that of a clock in the laboratory), second with respect to $\frac{1}{c}$ (post-Newtonian expansion). I consider the one-particle sector of a massive spinor field in QFT, here described effectively by a classical field. The formal expansion in powers of $\frac{1}{c}$ yields a systematic and complete generation of GR corrections for quantum systems. I find new terms that were overlooked in the literature at order $\frac{1}{c^2}$ and extended the level of approximation to the next order. These findings are significant for a consistent inclusion of gravity corrections in the description of quantum experiments of corresponding sensitivities, and also for testing aspects of GR, like the EEP, in the quantum realm.

\chapter*{Acknowledgement}

 Foremost, I would like to express my sincere gratitude to my supervisors Prof.\ Dr.\ Domenico Giulini and Dr.\ Philip K.\ Schwartz. Their immense knowledge, patience and enthusiasm was invaluable for me throughout my Master's studies and research including this very thesis. I would also like to express special gratitude to Prof.\ Dr.\ Klemens Hammerer for his willingness to be my second examiner.   

I also would like to thank T.\ Rick Perche and Jonas Neuser for a fruitful discussion about my results and my analysis of their work after I first uploaded this thesis to the arXiv, which has led to this updated version.

I would like to thank my two friends Bardia H. Fahim and Veljko Simovic for their constant supports and countless fruitful discussions on Physics, Mathematics and Philosophy. Specially, I thank Saina Barani for enduring with me in all the ups and downs I faced. I also want to thank my fellow students Luise Kranzhoff and Reejula Roy for their interest in my work and their questions that helped me understand my work from different perspectives. 

I am grateful to Rachel Sowerby and Lauren Muir for proofreading and checking this thesis grammatically.

Last but not the least, I would like to thank my family, my parents, my brother and my uncle and his wife for giving me the emotional support, financial support and for standing behind me in all of these years.

I dedicate this thesis to my cousin Nassim who recently discovered that she has
blood cancer. I am sure she will defeat it.

\tableofcontents 
\enlargethispage{\baselineskip}

\mainmatter

\chapter{Introduction} 

\epigraph{Algebra is the offer made by the devil to the mathematician. The devil said: \enquote{I will give you this powerful machine, it will answer any question you like. All you need to do is give me your soul: give up geometry and you will have this marvellous machine.}}{Michael Atiyah}

The primary aim of this thesis is to probe the post-Newtonian correction of the Dirac equation in the presence of the gravitational field. My thesis can be understood as an extension of work done by Schwartz \& Giulini \cite{schwartz19:AiG,schwartz19:PNSE}. I expand on Schwartz \& Giulini's work by adding the notion of spin which has led me to use the Dirac equation instead of the Klein-Gordon equation.

Considering gravity being described by GR, I use the Dirac equation because it avoids the shortcomings of the Schrödinger equation and grants us some important additional benefits.
The use of the Schrödinger equation is problematic because
the Schrödinger equation is:
(1) a non-relativistic equation; (2) incompatible with GR; and (3) a first derivative of time and second derivative of space which make it much harder to work with than the Dirac equation.
Using the Dirac equation is beneficial because the Dirac equation:
(1) has a clear geometrical description based on the concept of Spinors which are well defined in the curved spacetime; (2) is a relativistic equation; and (3) has a simpler differential structure in comparison to the Schrödinger equation.

My theoretical motivation is twofold: firstly to grasp a clear understanding of the complicated interaction of gravity with quantum particles in a curved background in QFT by considering their classical (i.e.\ `non-relativistic', but still quantum) limits. Secondly, to fill the gaps linking the description of spinor fields within classical Newtonian gravity to their description in relativistic GR, by providing a rigorous geometrical description for the first one.

As a prerequisite to examining the post-Newtonian correction, I expand the Dirac equation in the Fermi Normal Coordinate (FNC). The FNC can be seen as a generalisation of the notion of a \enquote{proper reference frame} to curved spacetime, defined with respect to a reference worldline, e.g.\ that of a clock in the laboratory.  Note that the worldline in question is arbitrary and not just geodesic. To perform the FNC expansion, the following two conditions (known as \enquote{weak gravity conditions}) need to be satisfied~\cite{doip:10.1063/1.1724316}:
\begin{itemize}
    \item The geodesics in the coordinate neighbourhood must not intersect i.e.\ the cur\-vature of the spacetime must be small compared to the size of the system.
    \item The spacelike hypersurfaces must not intersect i.e.\ the curvature of the worldline must be small compared to the size of the system.
\end{itemize}

With the FNC expansion in place, I turn to post-Newtonian framework.
The post-Newtonian framework replaces the abstract notion of Spinors with the far more palpable notion of wavefunction. The post-Newtonian framework transforms the Dirac equation to the Pauli equation which is equivalent to the modified Schrödinger equation that includes spin effects. The post-Newtonian framework has recently received a lot of attention. It has been applied in variety of areas ranging from experimental quantum optics to theoretical quantum gravity \cite{PhysRevD.101.045017,PhysRevA.103.013703}. The post-Newtonian framework is popular because it bridges the gap between the Newtonian physics and special relativistic theories. However, it is crucial to note that there is no consensus on what constitutes a post-Newtonian framework. While, among various methods, some insist on mass and energy as grounding parameters, others rely on Foldy--Wouthuysen transformation (FW-transformation) \cite{PhysRev.78.29}. Following Giulini \& Großardt \cite{Giulini_2012}, I take that the post-Newtonian framework should be grounded in the \enquote{formal c-power expansion}.

I contribute to what has been previously described by finding a way to systematically include GR correction terms in quantum systems, deriving new terms that were neglected in the literature \cite{perche21} and extending the level of approximation \cite{PhysRevLett.44.1559,PhysRevD.22.1922,perche21,Ito:2020xvp}. My findings are significant to experimental physics because they suggest that quantum particles can be used to test and measure gravity as described by GR. In addition, my results can be seen as the increase of measurement sensitivity of quantum experiments by including GR correction terms. One of the most important applications of my calculation is in the study of gravitational effects on $g$-factor measurement which has been already considered in \cite{10.1093/ptep/pty066,Visser2018PostNewtonianPP,Nikoli2018CanEM,Venhoek2018AnalyzingM,L_szl__2018,PhysRevA.98.032508,PhysRevD.100.064029}. However, the mentioned literature is neither complete nor systematic. My findings can also be used as a grounding for atomic interferometry observations and detection of the possible quantum gravity effects \cite{PhysRevLett.114.013001,PhysRevLett.118.183602,PhysRevLett.120.183604,PhysRevD.101.125018}. \footnote{For more inclusive list of applications see chapter \ref{chap:concl}.}

This thesis is organised as follows: The second chapter begins with the conventions and assumptions which are followed by mathematical and physical background needed to comprehend this thesis as intended. In chapter three, I expand the Dirac equation in the FNC in the extended order of approximation. This expansion not only illustrates the geometrical description of Spinors in curved spacetime, but it also provides us with the definition of time as it is measured in a laboratory setting. In chapter four, I calculate the post-Newtonian limit of the expanded Dirac equation in the FNC and the subsequent modifications to the Pauli equation. Finally, in chapter five, I compare my results with two papers \cite{perche21,Ito:2020xvp} that were published in 2021 and I argue that different post-Newtonian approaches lead to different results. My thesis concludes by claiming that the formal expansion in powers of $c$ seems to be more systematic in comparison to the other methods that are used in the literature.

This thesis contains 5 appendices, regarding explicit computations. We calculate the inverse of the metric which is expanded in the FNC in the appendix \ref{appendix A} and the corresponding Christoffel symbols in the appendix \ref{appendix B}. Then using the results in the previous appendices, we calculate the connection one-forms in appendix \ref{appendix C} and finally, the spin connection in appendix \ref{appendix D}. There are some computational tricks provided in appendix \ref{appendix E} which we will use to simplify our resulting Pauli Hamiltonian.

\chapter{Mathematical and physical background} \label{chap:background}

\section{Conventions and Assumptions}
 
Throughout this thesis, the following are presumed:
\begin{itemize}
\item  $\hbar=1$

\item The metric signature is `mostly plus', such that the Minkowski metric's components are $\eta_{IJ} = \mathrm{diag}(-1,+1,+1,+1)$

\item Inner product for two 3-dimensional vectors are shown as: $a^ib_i:=a\cdot b$, where $i=1,2,3$.

\item Vector products for two 3-dimensional vectors are shown as: $(a\times b)^i = \tensor{\epsilon}{^{i}_{jk}} a^j b^k$, where $i$,$j$ and $k$ run from $1$ to $3$. 

\item The vector arrows are dropped. Such that, terms such as  $\Vec{a} \cdot \Vec{b}$ and $(\Vec{a}\times \Vec{b})^i$ are written as $(a\cdot b)$ and $(a\times b)^i$ respectively.

\item When expanding any quantity $X$ in powers of $c^{-1}$, the term of order $n$, \emph{including the prefactor $c^{-n}$}, will be denoted by $X(n)$; i.e.\ we have
\begin{subequations} \label{expand_convention}
\begin{equation}
    X = X(0) + X(1) + X(2) + \dots
\end{equation}
with
\begin{equation}
    X(0) = O(c^0), \quad X(1) = O(c^{-1}), \quad \dots
\end{equation}
\end{subequations}
For example, this convention will be applied to Hamiltonians $H = H(0) + H(1) + \dots$ and fields $\psi = \psi(0) + \psi(1) + \dots$.

\item Unless stated otherwise, the symbols $\{$ and $\}$ are only brackets and detached from their use as anti-commutators. 

\item English indices such as $I=(0, \mathrm{i})$  are to be distinguished from their Greek counterparts such as $\mu=(s,i)$.
The former will be used for components with respect to tetrads (orthonormal frames), while the latter is for spacetime coordinate components.

\item The curvature tensor is assumed to be of the order of $c^{-2}$.

\item The timelike coordinate $s$ has the dimension of length and can be related to the proper time as $s=c\tau$.

\item The time derivative of acceleration, curvature and rotation are neglected.

\item By using the term `Fermi normal coordinate', we mean the \enquote{generalised Fermi normal coordinate}, in which the the rotation of the frame is also assumed. 

\item The order of approximation for Fermi Normal Coordinate expansion is up to and including $O(x^2)$.

\item The order of approximation for the post-Newtonian expansion is up to and including $O(c^{-2})$.
\end{itemize}

\section{The Dirac equation in flat and curved spacetime}

\subsection{The Dirac equation in flat spacetime}

Among the various definitions of Spinors, this thesis adheres to the following: Objects which transform under the action of the Lorentz group according to the $(\frac{1}{2},0)\oplus(0,\frac{1}{2})$ representation of the Lorentz algebra are called Spinors.
That means that Spinors can be treated as an element of a four-dimensional complex vector space:
\begin{align}
    \psi=\begin{pmatrix} \psi_1 \\ \psi_2 \\ \psi_3 \\ \psi_4
    \end{pmatrix}
\end{align}
wherein, each component is a complex function in the spacetime. This representation can be reduced to a two-component ($\psi_A$ and $\psi_B$) representation, both of which are a two-dimensional complex vector:
\begin{align} \label{yx}
      \psi= \begin{pmatrix} \psi_A \\ \psi_B \end{pmatrix}
\end{align}
In this representation, $\psi_A$ is interpreted as the state of positive frequencies (particles) and $\psi_B$ as the state of negative frequencies (anti-particles). 

The Dirac equation is meant to describe the spin-½ massive particles. It is a relativistic equation meaning that, it respects the principles of the theory of Special Relativity.
Originally, the Dirac equation was written in the flat spacetime background. However, there exist non-trivial ways of writing it in the curved background spacetime. This equation is usually expressed in its simplest form \cite{Shankar:102017}: 
\begin{align}
 (i \gamma^\mu\partial_\mu -mc)\psi =0   
\end{align}
where, $\gamma^\mu$ are so-called Gamma matrices:
\begin{align} \label{xy}
    \gamma^0=\begin{pmatrix} \mathbb{1} & 0 \\
    0& -\mathbb{1}
    \end{pmatrix} , \qquad   \gamma^i=\begin{pmatrix} 0 & \sigma^i \\
    -\sigma^i & 0
    \end{pmatrix}
\end{align}
while, $m$ is the mass and $c$ is the speed of light .

Gamma matrices are generators of the Clifford algebra. Their anti-commutation relations are as follows:\footnote{Note that in order to use the standard convention for gamma matrices, we add an extra minus sign to the Clifford algebra anti-commutation relations. This is due to the mostly-plus convention we have for the Minkowski spacetime.}
\begin{align}
    \{\gamma^\mu ,\gamma^\nu \}=-2\eta^{\mu \nu}\mathbb{1}_4
\end{align}
where, $\eta_{\mu \nu}=(-1,1,1,1)$ represent the Minkowski spacetime,
and the so-called Pauli matrices are:
\begin{align}
    \sigma^1=\begin{pmatrix} 0 & 1 \\
    1 & 0
    \end{pmatrix} ,\qquad
     \sigma^2=\begin{pmatrix} 0 & -i \\
    i & 0
    \end{pmatrix} , \qquad
     \sigma^3=\begin{pmatrix} 1& 0 \\
    0 & -1
    \end{pmatrix}
\end{align}
for which, the following (anti-)commutation relations are considered:
\begin{align}
 &[\sigma^i,\sigma^j]=2i\tensor{\epsilon}{^{ij}_k}\sigma^k \nonumber\\
 &\{\sigma^i , \sigma^j\}= 2\delta^{ij}\mathbb{1}
\end{align}
from which, a very useful relation can be derived:
\begin{align}\label{2.8}
    \sigma^i \sigma^j=\delta^{ij}+i\tensor{\epsilon}{^{ij}_k}\sigma^k
\end{align}
Furthermore, one could easily include the electromagnetism interaction into the Dirac equation by changing the partial derivative into a covariant derivative including the electromagnetic four-potential:
\begin{align}
 (i\gamma^\mu D_\mu -mc)\psi =0   ; \qquad D_\mu=\partial_\mu -iq A_\mu
\end{align}

\subsection{Spinors in curved spacetime}

In order to generalise the Dirac equation in curved spacetime, the flat version of the equation needs to be modified in a way that it respects the principles of the theory of General Relativity \cite{Wald:106274,Sakurai:1167961}.
In order to do so, we need to use the formalism of Spinors in four-dimensional Lorentzian manifolds. It means we need to provide the description in which Spinors are defined within the context of curved spacetime . This can be done by Spinor bundle formalism. That alludes to define Spinor bundle as an associated bundle to a $SO(1,3)$ principal bundle. As a matter of fact, $SL(2,\mathbb{C})$ is the double cover of $SO(1,3)$ and it allows us to formulate the Spinor bundle from the associated bundle and find the natural connection derived from it.
More explicitly, using the tetrad (vierbein) formalism, we can set $e_I=(e_0, e_\mathrm{i})$ to be an orthonormal frame. That equates to:
\begin{align}
    e^\mu_I e^\nu_J g_{\mu \nu}=\eta_{IJ}
\end{align}
We can define the connection 1-form associated with this frame as:
\begin{align}
    \nabla_\mu e_I = \tensor{w}{_\mu^J_I}e_J
\end{align}
Now we can introduce another frame equivalent to Dirac basis for Spinor in flat spacetime: $E_a$  where; $a=1,2,3,4$ on Spinor bundle. In other words, we demand this new frame to preserve the form of gamma matrices as they were introduced earlier. 
Therefore we can introduce the covariant derivative as follows:
\begin{align}
\nabla_\mu \psi^a = \partial_\mu \psi^a+\tensor{\Gamma}{_\mu^a_b}\psi^b    
\end{align}
We know that we can also express the spin connections in terms of the group generator and the connection one-form \cite{2019}:
\begin{align}
 \tensor{\Gamma}{_\mu^a_b}\psi^b =-\frac{1}{2}w_{\mu IJ} \tensor{(S^{IJ})}{^a_b}\psi^b
\end{align}
where:
\begin{align} \label{plpl}
     S^{IJ}=\frac{1}{4}[\gamma^I,\gamma^J]
\end{align}
is the generator of the group $SL(2,\mathbb{C})$ action on the Spinor bundle and $w_\mu$ is the connection one-form. Note that in order to define the associated Spinor bundle, we demand our orientable manifold to have the spin structure. On top of these conditions, we will also demand that there exists a globally defined orthonormal frame over our manifold so that Geroch’s theorem ensures that a spinor bundle can be built \cite{stewart_1991,penrose_rindler_1984}.

\section{Newtonian limit: Dirac to Pauli in flat spacetime}

There are many approaches to study the post-Newtonian limit. FW-transformation,  $\frac{1}{m}$ expansion and formal $c$ power expansion are examples of such. In the thesis, I will use the formal $c$ power expansion and then in chapter \ref{chap5}, we will compare it to other methods to see the differences. 

By formal expansion, it is meant to formulate the expansion as the formal limit of $c\rightarrow\infty$, where, $c$ is the speed of light. To implement this limit, we will expand Spinors in the Dirac equation as formal power series in the parameter $c^{-1}$ - or, more precisely, formal Laurent series, since we will need negative orders of $c$. 

Analytically speaking, a Taylor expansion in a dimensionful parameter like $c$ is incorrect. Therefore we need to make it dimensionless as a small-parameter approximation. 
This means that the corresponding small parameter has to be chosen as e.g., the ratio of some typical velocity of the system under consideration to the speed of light. In this thesis however, we will drop this issue and just expand in $c^{-1}$ as a formal parameter. 

More strictly speaking, we are to deal with tensor field that takes values in the field of formal Laurent series $\mathbb{R}{((c^{-1}))}$ instead of the real numbers. It does not cause any problem because, we define differentiation of series-valued tensor field order by order, and demand that equations be satisfied order by order \cite{Philip_Newton}.

Let us see how to derive Pauli equation as the post-Newtonian limit of Dirac in flat spacetime:
\begin{align}
    (i\gamma^\mu \partial_\mu -mc)\psi=0
\end{align}
Using \eqref{yx} , \eqref{xy} and separating the spatial and temporal part will result in:
\begin{align} \label{ttt}
    &(i\partial_t-mc^2)\psi_A =-i c(\sigma \cdot \partial)\psi_B \nonumber\\ 
    &(i\partial_t+mc^2)\psi_B =-i c(\sigma \cdot \partial)\psi_A
\end{align}
Now we separate the phase factor:
\begin{align} \label{2.18}
    \psi=e^{ic^2 s} \Tilde{\psi}\; ,\quad \Tilde{\psi}= O(c^0) 
\end{align}
where, $s$ is an arbitrary function of space and time. 
Inserting it to \eqref{ttt}:
\begin{subequations}
\begin{align} 
    &(-c^2\Dot{s}+i  \partial_t-mc^2)\Tilde{\psi}_A =(c^3 (\sigma\cdot\partial s) - ic\sigma \cdot \partial)\Tilde{\psi}_B \qquad \label{eq:pauli_expand_1} \\
    &(-c^2\Dot{s}+i \partial_t+mc^2)\Tilde{\psi}_B =(c^3 (\sigma\cdot\partial s) -i c\sigma \cdot \partial)\Tilde{\psi}_A\qquad \label{eq:pauli_expand_2}
\end{align}
\end{subequations}
where, $\Dot{s}=\partial_ts$ and $s^\prime$ is the spatial derivative of $s$.
Now we are ready to expand our equations in the powers of $c$. That means that we expand $\tilde\psi_A$ and $\tilde\psi_B$ as in \eqref{expand_convention}.

Then we can look at the different orders of $c$:

\eqref{eq:pauli_expand_1} at $c^3$:
\begin{align}\label{11}
    (\sigma\cdot\partial s)\Tilde{\psi}_B(0)=0
\end{align}

\eqref{eq:pauli_expand_2} at $c^3$:
\begin{align}\label{12}
  (\sigma\cdot\partial s)\Tilde{\psi}_A(0)=0
\end{align}
Therefore in order to have a non-zero field we conclude:
\begin{align}
    \partial_i s=0
\end{align}
This implies that $s$ depends only on time.

\eqref{eq:pauli_expand_1} at $c^2$: 
\begin{align}
   (\Dot{s}+m)\Tilde{\psi}_A(0)=- (\sigma \cdot \partial s)\Tilde{\psi}_B(1)
\end{align}

\eqref{eq:pauli_expand_2} at $c^2$:
\begin{align}
   (-\Dot{s}+m)\Tilde{\psi}_B(0)=(\sigma \cdot \partial s)\Tilde{\psi}_A(1)
\end{align}

Knowing that $s$ is not function of space, the right hand side is vanished:

\eqref{eq:pauli_expand_1} at $c^2$:
\begin{align}\label{13}
   (\Dot{s}+m)\Tilde{\psi}_A(0)=0
\end{align}

\eqref{eq:pauli_expand_2} at $c^2$:
\begin{align}\label{14}
   (-\Dot{s}+m)\Tilde{\psi}_B(0)=0
\end{align}
Again, as we want to have a non-zero field, we conclude:
\begin{align}
   &(\Dot{s}+m)=0 \nonumber\\
   &(-\Dot{s}+m)=0 
\end{align}
Therefore we can derive the explicit function of $s$ as follows:
\begin{align}
    s=\pm mt +K
\end{align}
where, we can choose $K=0$ because it is a phase in the exponential function. In addition, we can ignore the $s=+mt$ case by being interested in positive energy solutions. Accordingly, \eqref{2.18} will turn into a strong ansatz:
\begin{align}
    \psi=e^{-imc^2t} \Tilde{\psi}
\end{align}
Therefore the equation \eqref{ttt} will turn to:
\begin{subequations}
\begin{align} 
    i\partial_t\Tilde{\psi}_A &=(-i c\sigma \cdot \partial)\Tilde{\psi}_B \qquad \label{eq1}\\
    (i\partial_t+2mc^2)\Tilde{\psi}_B &=(-i c\sigma \cdot \partial)\Tilde{\psi}_A\qquad \label{eq2}
\end{align}
\end{subequations}
Now that we have fixed our ansatz, we can confidently expand our Dirac equation. For the order of $c^3$ (\eqref{11} and \eqref{12}) due to $s^\prime=0$, there is no equation. However for the next order, namely $c^2$, our equations \eqref{13} and \eqref{14} will result in:
\begin{align} \label{ooo}
    \Tilde{\psi}_B(0)=0
\end{align}
which confirms our ansatz. With this information, we can now study next orders and solve the coupled equations for the positive frequencies ($\Tilde{\psi}_A$):

\eqref{eq1} at $c^1$:
\begin{align}\label{asd}
0=(-i c\sigma \cdot \partial)\Tilde{\psi}_B(0)
\end{align}

\eqref{eq2} at $c^1$:
\begin{align} \label{ash}
\Tilde{\psi}_B(1)=-\frac{i}{2mc}(\sigma \cdot \partial)\Tilde{\psi}_A(0)   
\end{align}
where, \eqref{asd} is trivial because of \eqref{ooo}. However, \eqref{ash} will be useful in the next order for finding an equation for $\psi_A(0)$:

\eqref{eq1} at $c^0$:
\begin{align}\label{lapp}
i\partial_t \Tilde{\psi}_A(0)=(-i c\sigma \cdot \partial)\Tilde{\psi}_B(1)
\end{align}

\eqref{eq2} at $c^0$:
\begin{align} \label{lax}
\Tilde{\psi}_B(2)=-\frac{i}{2mc}(\sigma \cdot \partial)\Tilde{\psi}_A(1)   
\end{align}
Using \eqref{lapp} and \eqref{ash}, we conclude:
\begin{align} \label{pauli}
    i\partial_t \Tilde{\psi}_A(0)=-\frac{1}{2m}(\sigma \cdot \partial)^2 \Tilde{\psi}_A(0)
\end{align}
Equation \eqref{pauli} is the Pauli equation. It is derived as the Newtonian limit ($c^0$ order) of the Dirac equation. We calculated it by expanding the Dirac equation in the powers of $c$. We can continue the procedure in order to see the post-Newtonian modification to the Pauli equation by looking at the higher orders of $c$:

\eqref{eq1} at $c^{-1}$:
\begin{align} \label{horn}
i\partial_t \Tilde{\psi}_A(1)=(-ic\sigma \cdot \partial)\Tilde{\psi}_B(2)
\end{align}
Using \eqref{horn} and \eqref{lax} we have again the Pauli equation, but this time for the $\Tilde{\psi}_A(1)$. 
That means there is no modifications to Pauli equation up to the first order of post-Newtonian approximation.
In the next order we have:

\eqref{eq2} at $c^{-1}$:
\begin{align} \label{aks}
i\partial_t\Tilde{\psi}_B(1)+2mc^2\Tilde{\psi}_B(3)=(-ic\sigma \cdot \partial)\Tilde{\psi}_A(2) 
\end{align}

\eqref{eq1} at $c^{-2}$:
\begin{align} \label{kos}
i\partial_t \Tilde{\psi}_A(2)=(-ic\sigma \cdot \partial)\Tilde{\psi}_B(3)
\end{align}
Replacing $\Tilde{\psi}_B(3)$ in \eqref{kos} by \eqref{aks}:
\begin{align}
    i\partial_t \Tilde{\psi}_A(2)=\frac{1}{2mc^2}(-ic\sigma \cdot \partial)\bigg\{ (-ic\sigma \cdot \partial)\Tilde{\psi}_A(2)-i\partial_t\Tilde{\psi}_B(1) \bigg\}
\end{align}
Now we use \eqref{ash} to replace $\Tilde{\psi}_B(1)$:
\begin{align}
    i\partial_t \Tilde{\psi}_A(2)=\frac{1}{2mc^2}(-ic\sigma \cdot \partial)\bigg\{ (-ic\sigma \cdot \partial)\Tilde{\psi}_A(2)-i\partial_t\{-\frac{i}{2mc}(\sigma \cdot \partial)\Tilde{\psi}_A(0)\}  \bigg\}
\end{align}
Simplifying it, we will end up:
\begin{align}
    i\partial_t \Tilde{\psi}_A(2)=-\frac{1}{2m}(\sigma \cdot \partial)^2 \Tilde{\psi}_A(2)-\frac{i}{4m^2c^2}(\sigma \cdot \partial)\partial_t(\sigma \cdot \partial)\Tilde{\psi}_A(0)
\end{align}
It can be concluded that in the $c^{-2}$ level of approximation, there are some modifications to Pauli equation. More explicitly, the second term of the above equation is considered to be the modification. Note that in order to include the external electromagnetic field interaction, one could replace the partial derivative $\partial$ by the covariant derivative $D_\mu=\partial_\mu-iqA_\mu$.
This method of relating the Dirac equation to the Pauli equation will be also applied in the curved spacetime context. But before that, we need another mathematical tool to be able to go to the curved spacetime scenario, namely, Fermi normal coordinates.

\section{Fermi normal coordinates}

Fermi normal coordinates, in short FNC,  are meant to be the generalisation of the inertial coordinates for  arbitrary timelike trajectories. These coordinates will be the second tool (first was post-Newtonian expansion) to reduce Dirac's Spinor formalism to an understandable experimental framework for arbitrary trajectories in curved spacetimes. As the post-Newtonian paradigm assumes \enquote{slow velocities}, metric expansion in the trajectory's neighbourhood will be done under the assumption of \enquote{weak gravity}.
The coordinates for general relativistic situations are related to observable quantities such as acceleration, rotation and distances. They were first formulated by M.Fermi in 1922. Then they were generalised by Manasse and Misner \cite{doip:10.1063/1.1724316} and later extended for the case of an arbitrary moving observer (i.e.\ rotation and acceleration) by Ni and Zimmermann \cite{q:PhysRevD.17.1473,Marzlin1994ThePM}.

Let us assume an arbitrary trajectory, to which, we ascribe a coordinate for a moving particle. The next step is to assume a laboratory around the particle on the trajectory. That means we are interested not just in the coordinate of the particle on the trajectory, but also the neighbourhood of that particle, or simply, the coordinate adapted to the observer in the laboratory.

Mathematically it means that we need to first define a co-moving orthonormal coordinates set for an arbitrary point in our trajectory. In order to do so, we will first set our timelike unit vector \cite{2011}. This can be easily done by assuming the velocity of the worldline, or more precisely, the normalised tangent vector to be the $e_0$:
\begin{align}
    e_0=\frac{\partial}{\partial\tau}
\end{align}
Now at each point of the worldline we construct a tetrad $e_I^\mu$ such that:
\begin{align} \label{98}
    e_0^\mu=(\frac{\partial}{\partial\tau})^\mu =\Dot{z}^\mu_0
\end{align}
where, $z^\mu(\tau)$ is the worldline of the observer.
Now according to \cite{Misner1973} we can define the generator of infinitesimal Lorentz transformation as follows:
\begin{align}
   \nabla_u e_I= \frac{De_I}{D\tau}=-\Omega.e_I
\end{align}
with
\begin{align}
\Omega^{\mu \nu}=a^\mu u^\nu -a^\nu u^\mu +u_I \omega_J \epsilon^{IJ\mu\nu}
\end{align}
where, $u=e_0$ is the four-velocity, $a^\mu$ is the four-acceleration and $\omega^\mu$ is the four-rotation of the observer. 

If $\omega$ is zero, the observer would be Fermi-Walker-transporting his tetrad (gyroscope-type transport). If both $a$ and $\omega$ were zero, he would be freely falling (geodesic motion) and would be parallel-transporting his tetrad ($
    \frac{De_I}{D\tau}=0$)
\cite{Misner1973}.
The last remark is that any normal tangent vector $v^\mu$ which is perpendicular to our timelike unit vector \eqref{98} can be written as \cite{Marzlin1994ThePM}:
\begin{align}
    v^\mu=e^\mu_i \alpha^i \qquad \text{where} \quad \sum_{i=1}^{3} (\alpha^i)^2=1
\end{align}

\begin{figure}
\begin{center}
\begin{tikzpicture}

\filldraw [black] (2,.5) circle (1pt);
\draw[gray] (2,.5)..controls(0,.3).. (-1,.5) node[left]{$x^\mu$} ;
\filldraw [gray] (-1,.5) circle (1pt)node[below=2pt] {\qquad  \qquad \qquad\qquad $x$};

\draw[thick] (2.2,0)..controls (2.4,-.7) and (1.8,-.3)..(2,-1.5);
\draw[very thick,-stealth] (1,-2)node[left] {$\tau$}--(1,-1);
\draw[thick] (2.2,0)--(2,.5);
\draw[thick] (2,.5)--(1.7,1.2);
\draw[thick] (2,-1.5)--(2.1,-1.9);
\draw [thick] (2.1,-1.9)--(2.3,-2.4);
\filldraw [black] (2.14,-2) circle (1pt);
\draw[red,thick,-stealth] (2.14,-2)--(1.8,-1)node[above] {$e^{\mu}_{(0)}$};
\draw[red,thick,-stealth] (2.14,-2)--(3,-2.5)node[above] {$e^{\mu}_{(2)}$};
\draw[red,thick,-stealth] (2.14,-2)--(2.7,-1)node[above] {$e^{\mu}_{(3)}$};
\draw[red,thick,-stealth] (2.14,-2)--(1.9,-2.8)node[below] {$e^{\mu}_{(1)}$};
\end{tikzpicture}
\end{center}
\caption{$z^{\mu}_{(\tau)}$ is the worldline of the observer and the adopted coordinates can be seen in red. One can see how to relate the neighbouring point $x^{\mu}$ to this adopted coordinates. Note that $x$ is the proper length of the geodesic connecting the worldline to the neighbouring point $x^{\mu}$. } \label{sddl}
\end{figure}
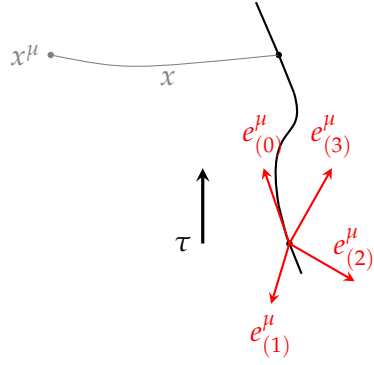

Now we can use the tetrad defined on the worldline to construct coordinates $x^\mu$ for any point in the neighbourhood of the worldline (see figure \ref{sddl}). We do so by connecting any points on the worldline to the neighbouring point by a geodesic starting at the point on the worldline with a tangent vector perpendicular to \eqref{98}. It will give us the advantage to describe all the neighbouring points by the proper time of the starting point in the worldline which we will call $\tau_0$ and three variables $\alpha^i$ and the length of the geodesic joining two points which, we call $x$.
 
Note that we should not confuse $x^\mu$ and proper length $x$. The latter one can be shown to be related to the spatial coordinate of the FNC by $x^i=x\alpha^i$.
That means we can expand the metric \cite{q:PhysRevD.17.1473} around this trajectory ($x^i=0$) by:
\begin{align}
    &g_{ss}= -1-2c^{-2}(a\cdot x)-c^{-4}(a\cdot x)^2 - R_{0l0m}x^lx^m +c^{-2}(\omega \times x)^2 + O(x^{3}) \\
    &g_{si}= c^{-1}(\omega \times x)_i - \frac{2}{3}R_{0lim}x^lx^m + O(x^{3})\\
    &g_{ij}= \delta_{ij}- \frac{1}{3}R_{iljm}x^lx^m+ O(x^{3})
\end{align}
where, $x$ can be also seen as the distance to the worldline. It is assumed that $x$ is small compared to the curvature radius. (weak gravity and weak inertial effects). $R_{ijkl}$ is the Riemann tensor, $a$ is acceleration of the observer on the worldline and $\omega$ is rotation of the frame. One should bear in mind that, the conditions for using the FNC can be summarised in the following two points \cite{Ito:2020xvp}:
\begin{itemize}
    \item The geodesics must not intersect i.e.\ The curvature of the spacetime must be small compared to the size of the system:
    \begin{align*}
        \tensor{R}{_{IJKL}} \cdot x^2 \ll 1 \Longleftrightarrow R_\text{system} \ll R_\text{spacetime} 
    \end{align*}
    \item The spacelike hypersurfaces must not intersect i.e.\ the curvature of the worldline must be small compared to the size of the system:
    \begin{align*}
        \frac{\omega}{c}\cdot x \ll 1 \Longleftrightarrow R_\text{system} \ll R_\text{angular velocity} 
\end{align*}
\begin{align*}
        \frac{a}{c^2} \cdot x \ll 1 \Longleftrightarrow R_\text{system} \ll R_\text{acceleration} 
    \end{align*}
\end{itemize}

One remark would be that the curvature tensor has a factor of $c^{-2}$ inside it. Therefore the metric can be seen as the Minkowski metric which is perturbed by terms of the different order of $c$.\footnote{This assumption that the curvature tensor has a factor of $c^{-2}$ inside it, only plays a role in the second phase of calculation where we expand the Dirac equation in terms of powers of $1/c$ to compute the post-Newtonian limits.}
Note that we expand the metric in terms of the FNC to the order of $x^2$ without being concern with the linearity of the acceleration, curvature and rotation.

It is noteworthy that in order to differ the frame indices from the ones for spacetime associated to Fermi normal coordinate indices, we will use the notation that Greek indices are mostly used to indicate the spacetime indices. They split into space and time components. The timelike coordinate $s$ has the dimension of length and can be related to the proper time as $s=c\tau$. The spatial components is indicated by italic $i$. The indices with respect to a chosen tetrad (orthonormal frame) are indicated by Latin indices, such that, $0$ for time coordinate and straight $\mathrm{i}$ for spatial components.
\begin{align}
    \mu=(s,i) \nonumber\\
    I=(0,\mathrm{i})
\end{align}
Sometimes throughout the calculation I might replace the frame indices with spacetime indices. That is the case only when those terms are to be calculated on the worldline itself and not the neighbourhood. In this case, it does not make a difference in which set of indices they are written.

\chapter{Expansion of the Dirac equation in Fermi normal coordinates}
\section{Dirac Hamiltonian}

In order to calculate the expanded Dirac equation in the FNC, we will first construct the Dirac Hamiltonian. To calculate the Hamiltonian, we remember the Dirac equation:
\begin{align}
    (i\gamma^\mu \nabla_\mu -mc)\psi=0
\end{align}
Splitting the index $\mu$ into its temporal and spatial components:
\begin{align}
    (i\gamma^s \nabla_s +i\gamma^i\nabla_i -mc)\psi=0
\end{align}
Inserting the covariant derivative and setting $s=c\tau$ for the partial time derivative will result in:
\begin{align}
    (i\gamma^s\partial_\tau +i\gamma^s c\Gamma_s +i\gamma^i c\partial_i + i\gamma^i c\Gamma_i -mc^2)\psi=0
\end{align}
Now we separate the time derivative in order to create a Schrödinger-like equation:
\begin{align}
    i\partial_\tau \psi=(-\gamma^s)^{-1}(i\gamma^i c\partial_i + i\gamma^i c\Gamma_i -mc^2)\psi -ic\Gamma_s \psi
\end{align}
which enables us to read-off the Hamiltonian. We call it Dirac Hamiltonian:
\begin{align} \label{3.35}
    H:=(g^{ss})^{-1}\gamma^s (i\gamma^ic \partial_i + i\gamma^i c\Gamma_i -mc^2)-ic\Gamma_s
\end{align}

\section{The spin connection in Fermi normal coordinates}

As we already saw, the general metric which is expanded in Fermi Normal Coordinate up to the order of $x^2$ is:
\begin{align}
    &g_{ss}= -1-2c^{-2}(a\cdot x)-c^{-4}(a\cdot x)^2 - R_{0l0m}x^lx^m + c^{-2}(\omega \times x)^2+ O(x^{3}) \\
    &g_{si}= c^{-1}(\omega \times x)_i - \frac{2}{3}R_{0lim}x^lx^m+ O(x^{3}) \\
    &g_{ij}= \delta_{ij}- \frac{1}{3}R_{iljm}x^lx^m+ O(x^{3})
\end{align}
From this, the inverse metric can be calculated (see appendix \ref{appendix A}):
\begin{align}
    &g^{ss}= -1+2c^{-2}(a\cdot x)-3c^{-4}(a\cdot x)^2 + R_{0l0m}x^lx^m + O(x^{3}) \\
    &g^{si}= c^{-1}(\omega \times x)^i - \frac{2}{3}\tensor{R}{_{0l}^i_m}x^lx^m -2c^{-3}(a\cdot x)(\omega \times x)^i+ O(x^{3}) \\
    &g^{ij}= \delta^{ij} + \frac{1}{3} \tensor{R}{^i_l^j_m}x^lx^m - c^{-2}(\omega \times x)^i (\omega \times x)^j+ O(x^{3})
\end{align}
Having the metric and its inverse, we can calculate the corresponding Christoffel symbols (see appendix \ref{appendix B}):
\begin{align}
    &\tensor{\Gamma}{^s_{ij}}= \frac{1}{3} \bigg\{(R_{0jil}+R_{0ijl})-2c^{-2}(a\cdot x)(R_{0jil} + R_{0ijl})-c^{-1}(\omega \times x)^n (R_{njil}+R_{nijl})\bigg\}x^l \\ \nonumber\\
    & \tensor{\Gamma}{^s_{si}} =c^{-2} a_i - c^{-4}a_i (a\cdot x)+ c^{-6} a_i (a\cdot x)^2 + R_{0i0l}x^l -2c^{-2}(a\cdot x)R_{0i0l}x^l \nonumber \\
    & \qquad- c^{-2}a_i R_{0l0m}x^lx^m -c^{-1}(\omega \times x)^n x^l R_{0lni} + \frac{2}{3}c^{-1} \tensor{R}{_{0l}^n_m}x^lx^m \epsilon_{ijn} \omega^j \\ \nonumber\\ \label{laklak}
    &\tensor{\Gamma}{^i_{ss}}=c^{-2}a^i + c^{-4}(a\cdot x)a^i + \tensor{R}{_0^i_{0l}}x^l - c^{-2}((\omega \times x)\times \omega)^i + \frac{c^{-2}}{3}\tensor{R}{^i_l^j_m}x^lx^m a_j  \nonumber\\
    & \qquad -c^{-4}a_j (\omega \times x)^i (\omega \times x)^j \\ \nonumber\\
    &\tensor{\Gamma}{^i_{sj}}=-c^{-3}(\omega \times x)^i a_j + c^{-5}a_j(a\cdot x)(\omega \times x)^i - c^{-1}(\omega \times x)^i R_{0j0l}x^l \nonumber \\ 
    & \qquad+ \frac{2}{3} c^{-2}a_j \tensor{R}{_{0l}^i_m}x^lx^m + c^{-1}\tensor{\epsilon}{^i_{pj}}\omega^p -\tensor{R}{_{0l}^i_j}x^l -\frac{1}{3}c^{-1}\epsilon_{npj}\omega^p \tensor{R}{^n_l^i_m}x^lx^m \\ \nonumber\\ \label{koso}
    &\tensor{\Gamma}{^s_{ss}}=c^{-3}(a.(\omega \times x))- c^{-5}(a.(\omega \times x))(a\cdot x)+c^{-1}R_{0n0l} x^l (\omega \times x)^n \nonumber\\
    & \qquad - \frac{2}{3}c^{-2}a_n \tensor{R}{_{0l} ^n _m}x^lx^m \\ \nonumber\\
    & \tensor{\Gamma}{^i_{jk}}= -\frac{1}{3}x^l c^{-1}(\omega \times x)^i (R_{0kjl} + R_{0jkl})-\frac{1}{3}x^l (\tensor{R}{^i_{kjl}} + \tensor{R}{^i_{jkl}})
\end{align}
Now we need to choose the FNC basis vectors:
\begin{align} \label{slm}
    &e^0_s=1+c^{-2}(a\cdot x)+\frac{1}{2}R_{0l0m}x^lx^m+ O(x^{3}) \nonumber \\
    &e^{\mathrm{i}}_s= - \frac{1}{2}\tensor{R}{^{\mathrm{i}}_{l0m}}x^lx^m +c^{-1}(\omega \times x)^{\mathrm{i}}+ O(x^{3}) \nonumber \\
    &e^0_j= \frac{1}{6}R_{0ljm}x^lx^m + O(x^{3})\nonumber \\
    &e^{\mathrm{i}}_j= \delta^{\mathrm{i}}_j -\frac{1}{6}\tensor{R}{^{\mathrm{i}}_{ljm}}x^lx^m + O(x^{3})
\end{align}
We can also calculate the dual basis vectors by:\footnote{Note that there is a difference between $j$ and $\mathrm{j}$ the former represent the spatial part of the spacetime and the latter represent the spatial part of the frame.}
\begin{align}
    &e_0^s=\eta_{00}(g^{ss}e^0_s+g^{sj}e^0_j) \nonumber\\
    &e_{\mathrm{i}}^s=\eta_{{\mathrm{ij}}}(g^{ss}e^{\mathrm{j}}_s+g^{si}e^{\mathrm{j}}_i) \nonumber\\
     &e_0^j=\eta_{00}(g^{js}e^0_s+g^{ji}e^0_i) \nonumber\\
    &e_{\mathrm{i}}^j=\eta_{{\mathrm{ij}}}(g^{js}e^{\mathrm{j}}_s+g^{ji}e^{\mathrm{j}}_i)
\end{align}

Therefore:
\begin{align} \label{salam}
    &e^s_0=1-c^{-2}(a\cdot x)+c^{-4}(a\cdot x)^2-\frac{1}{2}R_{0l0m}x^lx^m \nonumber \\
    &e^s_{\mathrm{i}}= - \frac{1}{6}\tensor{R}{_{\mathrm{i}}_{l0m}}x^lx^m  \nonumber \\
    &e^j_0= -c^{-1}(\omega \times x)^j +c^{-3}(a\cdot x)(\omega \times x)^j +\frac{1}{2}\tensor{R}{_{0l}^j_m}x^lx^m \nonumber \\
    &e^j_{\mathrm{i}}= \delta^j_{\mathrm{i}} +\frac{1}{6}\tensor{R}{_{\mathrm{i}}_l^j_m}x^lx^m 
\end{align}
Still requiring clarification is why did we chose this specific set of basis vectors. The reasoning will be provided in section \ref{okas} of chapter \ref{chap5} in which my calculation are compared with Ito \cite{Ito:2020xvp}.

Now we have the required information in order to calculate the connection one forms:
\begin{align}
    \tensor{w}{_\mu ^I_J}= \tensor{\Gamma}{^\alpha_{\mu \nu}}e^I_\alpha e^\nu_J -e^\nu_J \partial_\mu e^I_\nu
\end{align}
As the final step we need to calculate the derivative of our basis vectors $\partial_\mu e^I_\nu$. We first separate it to spatial and temporal parts for $\nu$, then we again separate each part to spatial and temporal parts for $I$. We also note that to simplify our calculations we ignore the time derivatives of acceleration, curvature and angular velocity. This can be physically interpreted as the metric being stationary up to our order of approximation.
Therefore the result is:
 \begin{align}
     &\partial_i e^0_s=c^{-2}a_i +R_{0i0l}x^l ;\;\;\;\;\;\;\; \mu=(s,i) \\
     &\partial_i e^{\mathrm{j}}_s=-\frac{1}{2}(\tensor{R}{^{\mathrm{j}}_{i0l}}+\tensor{R}{^{\mathrm{j}}_{l0i}})x^l +c^{-1}\epsilon^{j}_{\;\;ki}\omega^k ;\;\;\;\;\;\;\; \mu=(s,i)\\
     &\partial_j e^0_i= \frac{1}{6}(R_{0jil}+R_{0lij})x^l ;\;\;\;\;\;\;\; \mu=(s,j)\\
     &\partial_p e^{\mathrm{j}}_i = -\frac{1}{6}(\tensor{R}{^{\mathrm{j}}_{pil}}+\tensor{R}{^{\mathrm{j}}_{lip}})x^l ;\;\;\;\;\;\;\; \mu=(s,p)
 \end{align}
Now we are ready to calculate the connection one-forms (calculations can be found in appendix \ref{appendix C}). The result is: 
\begin{align}
    &\tensor{w}{_i^0_0}=\frac{c^{-1}}{2}x^l (\omega \times x)^k R_{0kli}+ \frac{1}{2}\epsilon_{ipn}\omega^p \tensor{R}{_{0l}^n_m}x^lx^m -c^{-2}(a\cdot x)R_{0i0l}x^l  \nonumber \\
    &\qquad-\frac{c^{-2}}{2}a_i R_{0l0m}x^lx^m \\
    & \tensor{w}{_s^{\mathrm{i}}_0}=c^{-2}a^i +c^{-1}(\omega \times x)^k \tensor{R}{_{0l}^i_k}x^l + \frac{c^{-1}}{2}\tensor{\epsilon}{^i_{pk}}\omega^p \tensor{R}{_{0l}^k_m}x^lx^m + \frac{c^{-2}}{6}\tensor{R}{^i_{ljm}}x^lx^ma^j \nonumber\\
    &\qquad -c^{-2} \tensor{R}{_0^i_{0l}}x^l(a\cdot x) +\tensor{R}{_0^i_{0l}}x^l - \frac{c^{-2}}{2}R_{0l0m}x^lx^m a^i \\
    &\tensor{w}{_i^{\mathrm{j}}_0}= \frac{c^{-1}}{6}\tensor{R}{^j_{lnm}}x^lx^m \tensor{\epsilon}{^n_{pi}}\omega^p + \frac{c^{-2}}{6}\tensor{R}{^j_{l0m}}x^lx^m a_i + \frac{1}{2}x^l\tensor{R}{^j_{0li}}- \frac{c^{-2}}{2}(a\cdot x)x^l \tensor{R}{^j_{0li}} \nonumber\\
    &\qquad +\frac{c^{-1}}{2}x^l (\omega \times x)^r \tensor{R}{^j_{ril}} \\ \label{dodo}
    &\tensor{w}{_s^0_{\mathrm{i}}}= -c^{-2}a_i - R_{0i0l}x^l +\frac{c^{-2}}{2}a_i R_{0l0m}x^lx^m +c^{-2}(a\cdot x)R_{0i0l}x^l + c^{-1}(\omega \times x)^nx^l R_{0lni} \nonumber \\
    & \qquad - \frac{c^{-2}}{6}a_j \tensor{R}{_{il}^j_m}x^lx^m + \frac{c^{-1}}{2}\tensor{\epsilon}{^n_{pi}}\omega^p R_{0lnm}x^lx^m \\\label{dododo}
    & \tensor{w}{_s^{\mathrm{i}}_{\mathrm{j}}}=\frac{c^{-1}}{6}\omega^p x^lx^m (\tensor{\epsilon}{^n_{pj}}R_{ilnm}+\tensor{\epsilon}{^n_{ip}}R_{jlnm}) +c^{-1}\tensor{\epsilon}{^i_{pj}}\omega^p -\tensor{R}{_{ol}^i_j}x^l \nonumber\\
    &\qquad  + \frac{c^{-2}}{6}x^lx^m (a_j \tensor{R}{^i_{l0m}}-a^i R_{jl0m}) \\
    & \tensor{w}{_k^{\mathrm{j}}_{\mathrm{i}}}= \frac{1}{2}\tensor{R}{^j_{ilk}}x^l \\
    &\tensor{w}{_\mu^0_0}=0 \\
    &\tensor{w}{_j^0_{\mathrm{i}}}= \frac{1}{2}R_{0ijl}x^l -\frac{c^{-2}}{3}x^l (a\cdot x)(R_{0jil} + R_{0ijl})-\frac{c^{-1}}{3}x^l (\omega \times x)^n (R_{njil} + R_{nijl})
\end{align}
Lastly, we are able to calculate the spin connection:
\begin{align}
    \Gamma_\mu =-\frac{1}{2} w_{\mu IJ}S^{IJ}
\end{align}
Separating it to spatial and temporal parts ($\mu=(s,f)$) we have (for details, see appendix \ref{appendix D}):
\begin{align} \label{3.29}
    &\Gamma_s= \frac{1}{2} \gamma^0 \gamma^{\mathrm{i}} \bigg\{ c^{-2}a_i +R_{0i0l}x^l -\frac{c^{-2}}{2}a_i R_{0l0m}x^lx^m -c^{-2}(a\cdot x)R_{0i0l}x^l -c^{-1}(\omega \times x)^n x^l R_{0lni} \nonumber \\
    &\qquad +\frac{c^{-2}}{6}a_j \tensor{R}{_{il}^j_m}x^lx^m -\frac{c^{-1}}{2} \tensor{\epsilon}{^n_{pi}}\omega^p R_{0lnm}x^lx^m \bigg\} \nonumber\\
    &\quad-\frac{1}{4}\gamma^{\mathrm{i}}\gamma^{\mathrm{j}} \bigg\{ \frac{c^{-1}}{6}\omega^p x^lx^m (\epsilon^{n}_{\;\;pj}R_{ilnm}+\epsilon^{\;\;\;n}_{ip}R_{jlnm}) +c^{-1}\epsilon_{ipj}\omega^p -R_{0lij}x^l \nonumber\\
    &\qquad +\frac{c^{-2}}{6}x^lx^m(a_j R_{il0m}-a_iR_{jl0m}) \bigg\}
\end{align}

 \begin{align} \label{3.30}
     &\Gamma_f = -\frac{1}{2} \gamma^{\mathrm{i}}\gamma^0 \bigg\{ \frac{c^{-1}}{6}R_{ilnm}x^lx^m \tensor{\epsilon}{^n_{pf}}\omega^p +\frac{c^{-2}}{6}R_{il0m}x^lx^m a_f + \frac{1}{2}x^l R_{i0lf}-\frac{c^{-2}}{2}(a\cdot x)x^l R_{i0lf} \nonumber \\
     &\qquad +\frac{c^{-1}}{2}x^l (\omega \times x)^r R_{irfl} \bigg\}- \frac{1}{4}\gamma^{\mathrm{i}}\gamma^{\mathrm{j}}\bigg\{ \frac{1}{2}R_{ijlf}x^l \bigg\}
 \end{align}

\section{The Dirac equation in Fermi normal coordinates} \label{oks}

There is a final step to be done before we are ready to expand the Dirac equation in the FNC. In order to relate the calculations we did for the expanded metric in the FNC, we need to expand the gamma matrices in terms of Fermi Normal Coordinates as well. That means:
\begin{align}
    &\gamma^s= e^s_0\gamma^0 +e^s_b \gamma^b ,\qquad I=(0,b) \\ 
    &\gamma^i= e^i_0 \gamma^0 + e^i_{\mathrm{j}}\gamma^{\mathrm{j}}, \qquad J=(0,{\mathrm{j}})
\end{align}
Therefore gamma matrices in the FNC can be read as follows:
\begin{align} \label{3.38}
    &\gamma^s= \bigg\{ 1-c^{-2}(a\cdot x) + c^{-4}(a\cdot x)^2 - \frac{1}{2}R_{0l0m}x^lx^m  \bigg\} \gamma^0 - \bigg\{ \frac{1}{6}R_{bl0m}x^lx^m \bigg\} \gamma^b \nonumber\\
    &\gamma^i= \bigg\{ -c^{-1}(\omega \times x)^i + c^{-3}(a\cdot x)(\omega \times x)^i +\frac{1}{2}\tensor{R}{_{0l}^i_m}x^lx^m \bigg\} \gamma^0 + \bigg\{\delta^i_j + \frac{1}{6}\tensor{R}{_{jl}^i_m}x^lx^m \bigg\} \gamma^{\mathrm{j}}
\end{align}

Now we are ready to calculate the Dirac Hamiltonian \eqref{3.35}. We have already calculated all the necessary terms in   \eqref{3.29} , \eqref{3.30} and \eqref{3.38}. The only term to be calculated is $(g^{ss})^{-1}$ which is calculated in appendix \ref{appendix A}.\footnote{Note that $g_{\mu\nu}^{-1}$ is not necessarily equal to $g^{\mu \nu}$. } It is, indeed, a colossal calculation and in order to avoid calculation mistakes, we divide it into parts and calculate it separately.

The first term we calculate is $(g^{ss})^{-1}\gamma^s$:
\begin{align} \label{wewe}
    (g^{ss})^{-1}\gamma^s&=\Bigg\{-1-2c^{-2}(a\cdot x)-c^{-4}(a\cdot x)^2-R_{0l0m}x^lx^m\Bigg\}\nonumber\\
    &\quad\Bigg\{ \bigg\{ 1-c^{-2}(a\cdot x) + c^{-4}(a\cdot x)^2 - \frac{1}{2}R_{0l0m}x^lx^m  \bigg\} \gamma^0 - \bigg\{ \frac{1}{6}R_{bl0m}x^lx^m \bigg\} \gamma^b\Bigg\} \nonumber\\
    &=\gamma^0\{-1-c^{-2}(a\cdot x)-\frac{1}{2}R_{0l0m}x^lx^m\}+\gamma^b\{\frac{1}{6}R_{bl0m}x^lx^m\}
\end{align}
Now we calculate $(g^{ss})^{-1}\gamma^s (i\gamma^ic \partial_i + i\gamma^i c\Gamma_i -mc^2)$ by multiplying \eqref{wewe} to each terms step by step. Therefore we first calculate $(g^{ss})^{-1}\gamma^s (i\gamma^ic \partial_i)$:
\begin{align} \label{3.40}
    (g^{ss})^{-1}\gamma^s (i\gamma^ic \partial_i)&=\Bigg\{\gamma^0\{-1-c^{-2}(a\cdot x)-\frac{1}{2}R_{0l0m}x^lx^m\}+\gamma^b\{\frac{1}{6}R_{bl0m}x^lx^m\}\Bigg\}\nonumber\\
    &\quad\Bigg\{ic\bigg\{ \{ -c^{-1}(\omega \times x)^i + c^{-3}(a\cdot x)(\omega \times x)^i +\frac{1}{2}\tensor{R}{_{0l}^i_m}x^lx^m \} \gamma^0 \nonumber\\
   & \qquad\qquad+ \{\delta^i_j + \frac{1}{6}\tensor{R}{_{jl}^i_m}x^lx^m \} \gamma^{\mathrm{j}}\bigg\}\partial_i\Bigg\} \nonumber\\
   &=\mathbb{1}\bigg\{i(\omega\times x)^i \partial_i-\frac{ic}{2}\tensor{R}{_{0l}^i_m}x^lx^m\partial_i\bigg\}-\gamma^0\gamma^j\bigg\{\delta^i_j\{ic\partial_i+ic^{-1}(a\cdot x)\partial_i\nonumber\\
   &\qquad+\frac{ic}{2}R_{0l0m}x^lx^m\partial_i\} +\frac{ic}{6}\tensor{R}{_{jl}^i_m}x^lx^m\partial_i\bigg\}+\gamma^b\gamma^j\bigg\{ \delta^i_j \{\frac{ic}{6}R_{bl0m}x^lx^m\partial_i\} \bigg\} 
\end{align}
Note that we used $\gamma^0\gamma^0=\mathbb{1}$ in the last step.

For the second term, namely $(g^{ss})^{-1}\gamma^s (i\gamma^i c\Gamma_i)$ we have to first replace the terms being careful with renaming the indices correctly:
\begin{align}
    (g^{ss})^{-1}\gamma^s (i\gamma^i c\Gamma_i)&=\Bigg\{\gamma^0\{-1-c^{-2}(a\cdot x)-\frac{1}{2}R_{0l0m}x^lx^m\}+\gamma^b\{\frac{1}{6}R_{bl0m}x^lx^m\}\Bigg\} \nonumber\\
    &\quad\Bigg\{ic \bigg\{\big\{ -c^{-1}(\omega \times x)^i + c^{-3}(a\cdot x)(\omega \times x)^i +\frac{1}{2}\tensor{R}{_{0l}^i_m}x^lx^m \big\} \gamma^0 +\nonumber\\
    &\qquad\big\{\delta^i_j + \frac{1}{6}\tensor{R}{_{jl}^i_m}x^lx^m \big\} \gamma^{\mathrm{j}}  \bigg\}\nonumber\\
    &\quad\bigg\{-\frac{1}{2} \gamma^{\mathrm{b}}\gamma^0 \big\{ \frac{c^{-1}}{6}R_{blnm}x^lx^m \tensor{\epsilon}{^n_{pi}}\omega^p +\frac{c^{-2}}{6}R_{bl0m}x^lx^m a_i + \frac{1}{2}x^l R_{b0li}\nonumber \\
     &\qquad-\frac{c^{-2}}{2}(a\cdot x)x^l R_{b0li}  +\frac{c^{-1}}{2}x^l (\omega \times x)^r R_{bril} \big\}- \frac{1}{4}\gamma^{\mathrm{b}}\gamma^{\mathrm{c}}\big\{ \frac{1}{2}R_{bcli}x^l \big\}
\bigg\} \Bigg\} \nonumber\\
\end{align}
where, for $\Gamma_f$ in \eqref{3.30} we have changed the index $i$ to $b$, $j$ to $c$ and $f$ to $i$ in order to carry on with the calculation.
Now we can easily multiply the terms and move forward in the calculation:
\begin{align}
    (g^{ss})^{-1}\gamma^s (i\gamma^i c\Gamma_i)&=ic\Bigg\{ \frac{1}{2}\gamma^b\gamma^0\big\{{-\frac{c^{-1}}{2}}(\omega \times x)^i  R_{b0li}x^l \big\}+\frac{1}{4}\gamma^b\gamma^c \big\{{-\frac{c^{-1}}{2}}(\omega \times x)^i R_{bcli}x^l \big\} \nonumber\\
    & \quad +\frac{1}{2}\gamma^0\gamma^j\gamma^b\gamma^0 \big\{ \delta^i_j\{\frac{c^{-1}}{6}R_{blnm}\tensor{\epsilon}{^n_{pi}}\omega^p x^lx^m +\frac{c^{-2}}{6}R_{bl0m}x^lx^m a_i+ \frac{1}{2} R_{b0li}x^l \nonumber\\
    & \qquad -\frac{c^{-2}}{2} (a\cdot x)x^l R_{b0li}+\frac{c^{-1}}{2}x^l (\omega \times x)^r R_{bril}+\frac{c^{-2}}{2}(a\cdot x)x^l R_{b0li}\big\} \nonumber\\
    & \quad + \frac{1}{4}\gamma^0 \gamma^j \gamma^b \gamma^c \big\{ \delta^i_j \{ \frac{1}{2}R_{bcli}x^l +\frac{c^{-2}}{2}(a\cdot x)R_{bcli}x^l\} \big\}\Bigg\} \nonumber\\
    &=\gamma^b\gamma^0\bigg\{-\frac{i}{4}x^l (\omega \times x)^i R_{b0li} \bigg\}+\gamma^b\gamma^c \bigg\{ -\frac{i}{8}x^l (\omega \times x)^i R_{bcli} \bigg\} \nonumber\\
    &+\gamma^j \gamma^b \bigg\{\frac{i}{12}R_{blnm}x^lx^m \tensor{\epsilon}{^n_{pj}}\omega^p +\frac{ic^{-1}}{12}R_{bl0m}x^lx^m a_j +\frac{ic}{4}x^l R_{b0lj}  \nonumber\\
    &\qquad -\frac{ic^{-1}}{4}(a\cdot x)x^l R_{b0lj} + \frac{i}{4}x^l (\omega \times x)^r R_{brjl} +\frac{ic^{-1}}{4}(a\cdot x)x^l R_{b0lj}\bigg\} \nonumber\\
    &+\gamma^0\gamma^j\gamma^b\gamma^c \bigg\{ \frac{ic}{8}R_{bclj}x^l +\frac{ic^{-1}}{8}(a\cdot x)R_{bclj}x^l \bigg\}
\end{align}
Note that we have used the fact that $\gamma^0\gamma^j\gamma^b\gamma^0=\gamma^j\gamma^b$ .
Now we can again rename $c$ to $j$ in order to further simplify the expression. We Note that $\gamma^0 \gamma^j=-\gamma^j \gamma^0$. Therefore we end up to:
\begin{align} \label{3.43}
    (g^{ss})^{-1}\gamma^s (i\gamma^i c\Gamma_i)&=
    \gamma^0\gamma^b\bigg\{\frac{i}{4}x^l (\omega \times x)^i R_{b0li} \bigg\} +\gamma^b \gamma^j \bigg\{-\frac{i}{8}x^l (\omega \times x)^i R_{bjli}\nonumber\\
    &\qquad +\frac{i}{12}R_{jlnm}x^lx^m \tensor{\epsilon}{^n_{pb}}\omega^p +\frac{ic^{-1}}{12}R_{jl0m}x^lx^m a_b+\frac{ic}{4}x^l R_{j0lb} -\nonumber\\
    &\qquad \frac{ic^{-1}}{4}(a\cdot x)x^l R_{j0lb} + \frac{i}{4}x^l (\omega \times x)^r R_{jrbl} +\frac{ic^{-1}}{4}(a\cdot x)x^l R_{j0lb}\bigg\} \nonumber\\
    &+\gamma^0\gamma^j\gamma^b\gamma^c \bigg\{ \frac{ic}{8}R_{bclj}x^l +\frac{ic^{-1}}{8}(a\cdot x)R_{bclj}x^l \bigg\}
\end{align}
Now the only term which needs to be simplified is $\gamma^0\gamma^j\gamma^b\gamma^c$. 
We know that we have the following relation for gamma matrices: 
\begin{align}
\gamma^I \gamma^J \gamma^K= -\eta^{IJ}\gamma^K - \eta^{JK} \gamma^I+\eta^{IK}\gamma^J-i\epsilon^{LIJK}\gamma_L \gamma^5
\end{align}
where, $\epsilon^{IJKL}$ is the Levi-Civita symbol in four dimensions and $\gamma^5=i\gamma^0 \gamma^1 \gamma^2 \gamma^3$ .
Therefore we can simplify the last line in \eqref{3.43} as follows:
\begin{align}\label{3.45}
    &\gamma^0\gamma^j\gamma^b\gamma^c \bigg\{ \frac{ic}{8}R_{bclj}x^l +\frac{ic^{-1}}{8}(a\cdot x)R_{bclj}x^l \bigg\}\nonumber\\
    &=\frac{ic}{8}\gamma^0\gamma^j\gamma^b\gamma^c R_{bclj}x^l \bigg\{1+c^{-2}(a\cdot x) \bigg\}\nonumber\\
    &=\frac{ic}{8}\gamma^0 R_{bclj}x^l (-\eta^{jb}\gamma^c - \eta^{bc} \gamma^j+\eta^{jc}\gamma^b-i\epsilon^{Ljbc}\gamma_L \gamma^5  ) \bigg\{ 1+c^{-2}(a\cdot x) \bigg\}\nonumber\\
    &=+\frac{ic}{8}\gamma^0 R_{jlbc}x^l(\eta^{jb}\gamma^c -\eta^{jc}\gamma^b)\bigg\{ 1+c^{-2}(a\cdot x) \bigg\}\nonumber\\
    &=-\frac{ic}{4}\gamma^0 R_{jlcb}x^l \eta^{jb}\gamma^c\bigg\{ 1+c^{-2}(a\cdot x) \bigg\} \nonumber\\
    &=-\frac{ic}{4}\gamma^0 \gamma^c (\eta^{JB}  R_{JlcB} -\eta^{00} R_{0lc0}) x^l\bigg\{ 1+c^{-2}(a\cdot x) \bigg\} \nonumber\\
    &=+\frac{ic}{4} \gamma^0 \gamma^c (R_{lc}+R_{0l0c})x^l\bigg\{ 1+c^{-2}(a\cdot x) \bigg\}
\end{align}
where, in the last two line of the above equations, we have used the fact that in order to take the trace of a tensor, we first need to include all the spatial and temporal indices. More explicitly:
\begin{align*}
    \eta^{jb}R_{jlbc} \neq R_{lc}
\end{align*}
Putting \eqref{3.45} back to \eqref{3.43}, we will end up:
\begin{align} \label{3.46}
    (g^{ss})^{-1}\gamma^s (i\gamma^i c\Gamma_i)&=
    \gamma^0\gamma^b\bigg\{\frac{i}{4}x^l (\omega \times x)^i R_{b0li} \bigg\} +\gamma^b \gamma^j \bigg\{-\frac{i}{8}x^l (\omega \times x)^i R_{bjli}\nonumber\\
    &\qquad +\frac{i}{12}R_{jlnm}x^lx^m \tensor{\epsilon}{^n_{pb}}\omega^p +\frac{ic^{-1}}{12}R_{jl0m}x^lx^m a_b+\frac{ic}{4}x^l R_{j0lb} -\nonumber\\
    &\qquad \frac{ic^{-1}}{4}(a\cdot x)x^l R_{j0lb} + \frac{i}{4}x^l (\omega \times x)^r R_{jrbl} +\frac{ic^{-1}}{4}(a\cdot x)x^l R_{j0lb}\bigg\} \nonumber\\
    &+\frac{ic}{4} \gamma^0 \gamma^c (R_{lc}+R_{0l0c})x^l\bigg\{ 1+c^{-2}(a\cdot x) \bigg\} \nonumber\\
    &=\gamma^b \gamma^j \bigg\{-\frac{i}{8}x^l (\omega \times x)^i R_{bjli}+\frac{i}{12}R_{jlnm}x^lx^m \tensor{\epsilon}{^n_{pb}}\omega^p +\nonumber\\
    &\qquad \qquad \frac{ic^{-1}}{12}R_{jl0m}x^lx^m a_b+ \frac{ic}{4}x^l R_{j0lb} -\frac{ic^{-1}}{4}(a\cdot x)x^l R_{j0lb} + \nonumber\\
    &\qquad \qquad \frac{i}{4}x^l (\omega \times x)^r R_{jrbl} +\frac{ic^{-1}}{4}(a\cdot x)x^l R_{j0lb}\bigg\} \nonumber\\
    & +\gamma^0 \gamma^c \Bigg\{+\frac{ic}{4}(R_{lc}+R_{0l0c})x^l\bigg\{ 1+c^{-2}(a\cdot x) \bigg\} +\bigg\{\frac{i}{4}x^l (\omega \times x)^i R_{c0li}\bigg\}\Bigg\}  
\end{align}
Finally, we calculate $(g^{ss})^{-1}\gamma^s (-mc^2)$ and $-ic\Gamma_s$ respectively as:
\begin{align}\label{3.47}
    (g^{ss})^{-1}\gamma^s (-mc^2)= \gamma^0\{mc^2+m(a\cdot x)+\frac{mc^2}{2}R_{0l0m}x^lx^m\}-\gamma^b\{\frac{mc^2}{6}R_{bl0m}x^lx^m\} 
\end{align}   

\begin{align}  \label{3.48}  
    -ic\Gamma_s&=-\frac{ic}{2} \gamma^0 \gamma^{\mathrm{i}} \bigg\{ c^{-2}a_i +R_{0i0l}x^l -\frac{c^{-2}}{2}a_i R_{0l0m}x^lx^m -c^{-2}(a\cdot x)R_{0i0l}x^l\nonumber \\
    &\qquad -c^{-1}(\omega \times x)^n x^l R_{0lni}  +\frac{c^{-2}}{6}a_j \tensor{R}{_{il}^j_m}x^lx^m -\frac{c^{-1}}{2} \tensor{\epsilon}{^n_{pi}}\omega^p R_{0lnm}x^lx^m \bigg\} \nonumber\\
    &\quad+\frac{ic}{4}\gamma^{\mathrm{i}}\gamma^{\mathrm{j}} \bigg\{ \frac{c^{-1}}{6}\omega^p x^lx^m (\epsilon_{npj}R_{ilnm}+\epsilon_{ipn}R_{jlnm}) +c^{-1}\epsilon_{ipj}\omega^p -R_{0lij}x^l \nonumber\\
    &\qquad +\frac{c^{-2}}{6}x^lx^m(a_j R_{il0m}-a_iR_{jl0m}) \bigg\}
\end{align}
Therefore we can insert \eqref{3.48} , \eqref{3.47} , \eqref{3.46} and \eqref{3.40} back into \eqref{3.35} and doing the simplifications, we have the Dirac Hamiltonian which is expanded in the Fermi Normal Coordinate as:
\begin{align}\label{eq:Dirac_FNC}
        H_\text{Dirac} &= \mathbb{1} \left\{ i(\omega \times x)^i D_i - \frac{ic}{2} \tensor{R}{_{0l}^i_m} x^l x^m D_i - q c A_0 \right\} \nonumber\\
    &\quad + \gamma^0 \left\{mc^2 + m(a\cdot x) + \frac{mc^2}{2} R_{0l0m} x^l x^m \right\} - \gamma^b \left\{ \frac{mc^2}{6}R_{bl0m}x^l x^m \right\}  \nonumber\\
    &\quad - \gamma^0 \gamma^j \bigg\{ icD_j + ic^{-1}(a\cdot x)D_j +\frac{ic}{2} R_{0l0m} x^l x^m D_j + \frac{ic}{6} \tensor{R}{_{jl}^i_m} x^l x^m D_i \nonumber\\
        &\qquad - \frac{i}{4} x^l (\omega \times x)^i R_{j0li} +\frac{ic^{-1}}{2}a_j +\frac{ic}{4} R_{0j0l} x^l -\frac{ic^{-1}}{2}a_j R_{0l0m}x^l x^m \nonumber\\
        &\qquad + \frac{ic^ {-1}}{4}(a\cdot x)R_{0j0l}x^l -\frac{i}{2} (\omega \times x)^i x^l R_{0lij}+ \frac{ic^ {-1}}{12} a_n R_{jl\;\;m}^{\;\;\;n}x^l x^m \nonumber\\
        &\qquad -\frac{i}{4}\epsilon^n_{\;\;pj}\omega^p R_{0lnm}x^l x^m - \frac{ic}{4} R_{lj}x^l -\frac{ic^{-1}}{4}(a\cdot x)R_{lj}x^l \bigg\} \nonumber\\
    &\quad+ \gamma^b \gamma^j \bigg\{ \frac{ic}{6} R_{bl0m}x^l x^m D_j +\frac{i}{4} \epsilon_{bpj} \omega^p + \frac{ic}{4}x^l R_{0bjl} \nonumber\\
        &\qquad + ix^l (\omega \times x)^r \left(\frac{1}{8}R_{jblr}+ \frac{1}{4} R_{jrbl}\right) + \frac{ic^{-1}}{24}x^l x^m (a_j R_{bl0m} + a_b R_{jl0m}) \nonumber\\
        &\qquad+ \frac{i}{24} \omega^p x^l x^m ( \tensor{\epsilon}{^n_{pb}} R_{jlnm} + \tensor{\epsilon}{^n_{pj}} R_{blnm} )  \bigg\}+ O(x^{3})
\end{align}
This is one of the main results of this thesis.

\chapter{Post-Newtonian expansion of the Dirac equation}

\section[Post-Newtonian expansion: Dirac to modified Pauli in curved spacetime]{Post-Newtonian expansion: Dirac to modified Pauli\\ in curved spacetime}

After expanding the Dirac equation in the FNC coordinate as the first phase of the calculation, we end up having the following Dirac Hamiltonian. We Note that it was calculated completely independent of the second upcoming phase: 
\begin{align} 
    H_\text{Dirac} &= \mathbb{1} \left\{ i(\omega \times x)^i D_i - \frac{ic}{2} \tensor{R}{_{0l}^i_m} x^l x^m D_i - q c A_0 \right\} \nonumber\\
    &\quad + \gamma^0 \left\{mc^2 + m(a\cdot x) + \frac{mc^2}{2} R_{0l0m} x^l x^m \right\} - \gamma^b \left\{ \frac{mc^2}{6}R_{bl0m}x^l x^m \right\}  \nonumber\\
    &\quad - \gamma^0 \gamma^j \bigg\{ icD_j + ic^{-1}(a\cdot x)D_j +\frac{ic}{2} R_{0l0m} x^l x^m D_j + \frac{ic}{6} \tensor{R}{_{jl}^i_m} x^l x^m D_i \nonumber\\
        &\qquad - \frac{i}{4} x^l (\omega \times x)^i R_{j0li} +\frac{ic^{-1}}{2}a_j +\frac{ic}{4} R_{0j0l} x^l -\frac{ic^{-1}}{2}a_j R_{0l0m}x^l x^m \nonumber\\
        &\qquad + \frac{ic^ {-1}}{4}(a\cdot x)R_{0j0l}x^l -\frac{i}{2} (\omega \times x)^i x^l R_{0lij}+ \frac{ic^ {-1}}{12} a_n R_{jl\;\;m}^{\;\;\;n}x^l x^m \nonumber\\
        &\qquad -\frac{i}{4}\epsilon^n_{\;\;pj}\omega^p R_{0lnm}x^l x^m - \frac{ic}{4} R_{lj}x^l -\frac{ic^{-1}}{4}(a\cdot x)R_{lj}x^l \bigg\} \nonumber\\
    &\quad+ \gamma^b \gamma^j \bigg\{ \frac{ic}{6} R_{bl0m}x^l x^m D_j +\frac{i}{4} \epsilon_{bpj} \omega^p + \frac{ic}{4}x^l R_{0bjl} + ix^l (\omega \times x)^r \left(\frac{1}{8}R_{jblr}+ \frac{1}{4} R_{jrbl}\right) \nonumber\\
        &\qquad + \frac{ic^{-1}}{24}x^l x^m (a_j R_{bl0m} + a_b R_{jl0m}) + \frac{i}{24} \omega^p x^l x^m ( \tensor{\epsilon}{^n_{pb}} R_{jlnm} + \tensor{\epsilon}{^n_{pj}} R_{blnm} )  \bigg\}
\end{align}
Now for taking the post-Newtonian limit as the second phase of calculation, we will do a similar process as in the flat spacetime case. First, we open the Dirac Hamiltonian for the $\psi_A$ and $\psi_B$ and end up having two equations for positive and negative frequencies. Then we apply the ansatz of positive frequencies by assuming $\psi= e^{-imc^2 t} \Tilde{\psi}$.
The result is the following two equations:
\begin{subequations}
 \begin{align}
   &\bigg\{ iD_\tau - m(a\cdot x) - \frac{mc^2}{2}R_{0l0m}x^l x^m -i(\omega \times x)^i D_i+ \frac{ic}{2}R_{ol\;\;\;m}^{\;\;\;i}x^l x^m D_i + \sigma^b \sigma^j \frac{ic}{6} R_{blom}x^l x^m D_j \nonumber\\
 &\quad+\frac{i}{4} \sigma^b \sigma^j \epsilon_{bpj}\omega^p+\frac{ic}{4} \sigma^b \sigma^j x^l R_{0bjl}  + i\sigma^b \sigma^j x^l (\omega \times x)^r \Big(\frac{1}{8}R_{jblr} + \frac{1}{4}R_{jrbl}\Big)\nonumber\\
 &\quad + \frac{ic^{-1}}{24}x^l x^m \sigma^b \sigma^j (a_j R_{bl0m} + a_b R_{jl0m}) + \frac{i}{24} \omega^p x^l x^m \sigma^b \sigma^j ( \epsilon^n_{\;\;pb} R_{jlnm} + \epsilon^n_{\;\;pj} R_{blnm} ) \bigg\} \Tilde{\psi}_A \nonumber\\
 &=\bigg\{-\frac{mc^2}{6}\sigma^b R_{blom}x^l x^m -\sigma^j icD_j -ic^{-1}(a\cdot x)\sigma^j D_j - \frac{ic}{2}\sigma^j R_{0l0m}x^l x^m D_j \nonumber\\
 &\qquad - \frac{ic}{6} \sigma^j R_{jl\;\;m}^{\;\;\;i} x^l x^m D_i + \frac{i}{4}\sigma^j x^l (\omega \times x)^i R_{joli}- \frac{ic^{-1}}{2} \sigma^j a_j - \frac{ic}{4}\sigma^j R_{0j0l}x^l \nonumber\\
 &\qquad+ \frac{ic^{-1}}{2}\sigma^j a_j R_{0l0m}x^l x^m - \frac{ic^{-1}}{4}\sigma^j (a\cdot x)R_{0j0l}x^l + \frac{i}{2}\sigma^j (\omega \times x)^i x^l R_{0lij}  \nonumber\\
 &\qquad -\frac{ic^{-1}}{12}\sigma^j a_n R_{jl\;\;m}^{\;\;\;n}x^l x^m +\frac{i}{4} \sigma^j \epsilon^n_{\;\;pj}\omega^p R_{0lnm}x^l x^m  + \frac{ic}{4} \sigma^j R_{lj}x^l + \frac{ic^{-1}}{4}\sigma^j (a\cdot x)R_{lj} x^l \bigg\} \Tilde{\psi}_B \label{eqone}
 \\[1em]
  &\bigg\{ iD_\tau + 2mc^2 +m(a\cdot x) + \frac{mc^2}{2}R_{0l0m}x^l x^m -i(\omega \times x)^i D_i+ \frac{ic}{2}R_{ol\;\;\;m}^{\;\;\;i}x^l x^m D_i\nonumber\\
 &\quad+ \sigma^b \sigma^j \frac{ic}{6} R_{blom}x^l x^m D_j + \frac{i}{4} \sigma^b \sigma^j \epsilon_{bpj}\omega^p +\frac{ic}{4} \sigma^b \sigma^j x^l R_{0bjl} \nonumber\\
 &\quad + i\sigma^b \sigma^j x^l (\omega \times x)^r \Big(\frac{1}{8}R_{jblr} + \frac{1}{4}R_{jrbl}\Big) + \frac{ic^{-1}}{24}x^l x^m \sigma^b \sigma^j (a_j R_{bl0m} + a_b R_{jl0m}) \nonumber\\
 &\quad+ \frac{i}{24} \omega^p x^l x^m \sigma^b \sigma^j ( \epsilon^n_{\;\;pb} R_{jlnm} + \epsilon^n_{\;\;pj} R_{blnm} ) \bigg\} \Tilde{\psi}_B  \nonumber\\
 &= \bigg\{\frac{mc^2}{6}\sigma^b R_{blom}x^l x^m -\sigma^j icD_j -ic^{-1}(a\cdot x)\sigma^j D_j - \frac{ic}{2}\sigma^j R_{0l0m}x^l x^m D_j - \frac{ic}{6} \sigma^j R_{jl\;\;m}^{\;\;\;i} x^l x^m D_i\nonumber\\
 &\qquad + \frac{i}{4}\sigma^j x^l (\omega \times x)^i R_{joli}- \frac{ic^{-1}}{2} \sigma^j a_j - \frac{ic}{4}\sigma^j R_{0j0l}x^l + \frac{ic^{-1}}{2}\sigma^j a_j R_{0l0m}x^l x^m \nonumber\\
 &\quad - \frac{ic^{-1}}{4}\sigma^j (a\cdot x)R_{0j0l}x^l + \frac{i}{2}\sigma^j (\omega \times x)^i x^l R_{0lij}  -\frac{ic^{-1}}{12}\sigma^j a_n R_{jl\;\;m}^{\;\;\;n}x^l x^m \nonumber\\
 &\quad +\frac{i}{4} \sigma^j \epsilon^n_{\;\;pj}\omega^p R_{0lnm}x^l x^m +\frac{ic}{4} \sigma^j R_{lj}x^l + \frac{ic^{-1}}{4}\sigma^j (a\cdot x)R_{lj} x^l  \bigg\} \Tilde{\psi}_A \label{eqtwo}
 \end{align}  
 \end{subequations}
Note that we have changed the partial derivative to the covariant derivative in order to include an external electromagnetic field. Now to see the post-Newtonian limit, we expand the $\Tilde{\psi} $  in c powers by noticing that the curvature tensor already has a factor of $c^{-2}$ inside. The result is: 
\begin{align}
   \Tilde{\psi} = \Tilde{\psi}(0) + \Tilde{\psi}(1) +  \Tilde{\psi}(2) +\dots
 \end{align}
\eqref{eqone} at $c^1$: 
\begin{align}\label{xxxx}
0 = -ic\sigma^j D_j \Tilde{\psi}_B(0)
\end{align} 
\eqref{eqtwo} at $c^2$:
\begin{align} \label{4.6}
\Tilde{\psi}_B(0)=0 
\end{align}
Therefore equation \eqref{xxxx} at $c^1$ becomes trivial.
We can carry on to the next order: \eqref{eqone} at $c^0$:
\begin{align} \label{xxo}
&\bigg\{i\partial_\tau+q c A_0 -m(a\cdot x) - \frac{mc^2}{2}R_{0l0m}x^l x^m - \nonumber \\
&\quad i (\omega \times x)^i D_i + \frac{i}{4} \sigma^b \sigma^j \epsilon_{bpj} \omega^p \bigg\} \Tilde{\psi}_A(0) =-\sigma^j ic D_j \Tilde{\psi}_B(1)
 \end{align}
\eqref{eqtwo} at $c^1$:
\begin{align}\label{4.8}
2mc^2 \Tilde{\psi}_B(1)= -\sigma^j icD_j \Tilde{\psi}_A(0) 
\end{align}
Putting the \eqref{4.8} into \eqref{xxo} at $c^0$:
\begin{align}
 &\bigg\{i\partial_\tau +q c A_0- m(a\cdot x)-\frac{mc^2}{2} R_\mathrm{0l0m}x^lx^m -\nonumber\\
 &i (\omega \times x)^i D_i + \frac{i}{4} \sigma^b \sigma^j \epsilon_{bpj} \omega^p \bigg\}\Tilde{\psi}_A(0)=\frac{-1}{2m} (\sigma \cdot D)^2 \Tilde{\psi}_A(0)
 \end{align}
So we can read off $H(0)$ by having $i\partial_t\Tilde{\psi}_A(0)=H(0)\Tilde{\psi}_A(0)$:

\begin{align} \label{hami0}
H(0)= -(\frac{1}{2m} ) (\sigma \cdot D)^2 +m(a\cdot x) + \frac{mc^2}{2}R_{0l0m}x^l x^m +i (\omega \times x)^i D_i - \frac{1}{2} \sigma_p \omega^p -q c A_0
\end{align}
where, for  $\frac{i}{4} \sigma^b \sigma^j \epsilon_{bpj} \omega^p$, we used:\begin{align*}
    \sigma^b \sigma^j = \delta^{bj} + \epsilon^{bj}_{\;\;\;k} \sigma^k
\end{align*}
in order to simplify it to $-\frac{1}{2}\sigma_p \omega^p$.
The next order then will be:
\eqref{eqone} at $c^{-1}$:
\begin{align} \label{pspsp}
&\bigg\{iD_\tau-m(a\cdot x)-\frac{mc^2}{2}R_{0l0m}x^l x^m -i (\omega \times x)^i D_i + \frac{1}{2} \sigma_p \omega^p \bigg\} \Tilde{\psi}_A(1) \nonumber\\
 &+ \bigg\{\frac{ic}{2}R_{0l\;\;m}^{\;\;\;i}x^l x^m D_i + \frac{ic}{6}\sigma^b\sigma^j R_{blom}x^lx^m D_j+ \frac{ic}{4}\sigma^b \sigma^j x^l R_{0bjl} \bigg\} \Tilde{\psi}_A(0) \nonumber\\
 &= \{-\sigma^j icD_j \} \Tilde{\psi}_B(2) - \{\frac{mc^2}{6} R_{bl0m}\sigma^b x^l x^m \} \Tilde{\psi}_B(1)
\end{align}
\eqref{eqtwo} at $c^0$:
\begin{align}
2mc^2 \Tilde{\psi}_B(2)=\{-\sigma^j icD_j\} \Tilde{\psi}_A(1) + \{ \frac{mc^2}{6}\sigma^b R_{blom}x^l x^m \} \Tilde{\psi}_A(0)
\end{align}
We can combine these two equations and have the expression for $\Tilde{\psi}_B(2)$ in terms of $\Tilde{\psi}_A(1)$ and $\Tilde{\psi}_A(0)$:
\begin{align} \label{4.14}
\Tilde{\psi}_B(2)= \frac{1}{2mc^2} \{ -\sigma^j icD_j\} \Tilde{\psi}_A(1) + \{\frac{1}{12} \sigma^b R_{bl0m}x^lx^m \} \Tilde{\psi}_A(0)
\end{align}
Putting \eqref{4.8} and \eqref{4.14} into \eqref{pspsp} at $c^{-1}$:
\begin{align}
& \bigg\{iD_\tau - m(a\cdot x)- \frac{mc^2}{2}R_{0l0m}x^l x^m -i (\omega \times x)^i D_i + \frac{1}{2} \sigma_p \omega^p \bigg\} \Tilde{\psi}_A(1)+\nonumber\\
 & \bigg\{\frac{ic}{2}R_{0l\;\;m}^{\;\;\;i}x^l x^m D_i + \frac{ic}{6}\sigma^b\sigma^j R_{blom}x^lx^m D_j+ \frac{ic}{4}\sigma^b \sigma^j x^l R_{0bjl} \bigg\} \Tilde{\psi}_A(0) \nonumber\\
 &=- \frac{1}{2m} (\sigma \cdot D)^2 \Tilde{\psi}_A(1) - \frac{1}{12}\bigg\{\sigma^j \sigma^b ic R_{bl0m}D_j(x^l x^m)\bigg\} \Tilde{\psi}_A(0) + \frac{ic}{12}R_{bl0m}\sigma^b \sigma^j x^l x^m D_j \Tilde{\psi}_A(0)
\end{align}
We can rename the operators acting on $\psi$s as follows: 
\begin{align}
i\partial_\tau \Tilde{\psi}_A(1)=H(1) \Tilde{\psi}_A(0)+ H(0) \Tilde{\psi}_A(1)
\end{align}
So we can read off the $H(1)$:
\begin{align} \label{4.17}
 H(1)= - \frac{2ic}{3} R_{0l\;\;m}^{\;\;\;i}x^l x^m D_i
 - \frac{ic}{4} \sigma^b \sigma^j x^l R_{0bjl} 
 - \frac{ic}{12} \sigma^j \sigma^b (R_{bj0l}+R_{bl0j})x^l  
\end{align}
Note that $H(0)$ is the same as the one calculated above. 
We finally need the next order of expansion in order to compute the Hamiltonian in $c^{-2}$ order. Therefore, \eqref{eqone} at $c^{-2}$ is:
\begin{align}\label{4.16p}
&\bigg\{iD_\tau -m(a\cdot x) -\frac{mc^2}{2}R_{0l0m}x^l x^m -i (\omega \times x)^i D_i + \frac{1}{2} \sigma_p \omega^p \bigg\} \Tilde{\psi}_A(2) \nonumber\\
 &\quad+ \bigg\{\frac{ic}{2}R_{0l\;\;m}^{\;\;\;i}x^l x^m D_i+\frac{ic}{6} \sigma^b \sigma^j R_{bl0m}x^l x^m D_j + \frac{ic}{4} \sigma^b \sigma^j x^l R_{0bjl} \bigg\} \Tilde{\psi}_A(1) \nonumber\\
 &\quad+ \bigg\{ i \sigma^b \sigma^j x^l (\omega \times x)^r (\frac{1}{8}R_{jblr} + \frac{1}{4}R_{jrbl}) + \frac{i}{24}\omega^p x^l x^m \sigma^b \sigma^j ( \epsilon^n_{\;\;pb}R_{jlnm}+\epsilon^n_{\;\;pj}R_{blnm} ) \bigg\}\Tilde{\psi}_A(0) \nonumber\\
 &
  =
  \bigg\{ - \sigma^j ic^{-1}(a\cdot x)D_j - \sigma^j \frac{ic}{2} R_{0l0m}x^l x^m D_j - \frac{ic}{6} \sigma^j R_{jl\;\;m}^{\;\;\;i}x^lx^m D_i -\frac{ic^{-1}}{2} \sigma^j a_j - \frac{ic}{4} \sigma^j R_{0j0l}x^l\nonumber \\
  & \quad+\frac{ic}{4} \sigma^j R_{lj}x^l \bigg\} \Tilde{\psi}_B(1)
+\{-\sigma^j icD_j\} \Tilde{\psi}_B(3)+ \{\frac{-mc^2}{6}R_{bl0m}x^lx^m \sigma^b \} \Tilde{\psi}_B(2)
\end{align}
\eqref{eqtwo} at $c^{-1}$:
\begin{align}
 &2mc^2 \Tilde{\psi}_B(3)+ \bigg\{iD_\tau + m(a\cdot x)+\frac{mc^2}{2}R_{0l0m}x^l x^m -i (\omega \times x)^i D_i + \frac{1}{2} \sigma_p \omega^p \bigg\} \Tilde{\psi}_B(1)\nonumber\\
 &=  \bigg\{ -\sigma^j ic^{-1} (a\cdot x) D_j - \sigma^j \frac{ic}{2}R_{0l0m}x^l x^m D_j - \frac{ic}{6} \sigma^j R_{jl\;\;m}^{\;\;\;i}x^l x^m D_i - \frac{ic^{-1}}{2}\sigma^j a_j - \frac{ic}{4}\sigma^j R_{0j0l}x^l \nonumber\\
 &\quad+\frac{ic}{4} \sigma^j R_{lj}x^l \bigg\} \Tilde{\psi}_A(0)+\{-\sigma^j icD_j\} \Tilde{\psi}_A(2) + \{ \frac{mc^2}{6} \sigma^b R_{bl0m}x^l x^m \} \Tilde{\psi}_A(1)
\end{align}
We can express the $\Tilde{\psi}_B(3)$ from the second equation in terms of $\Tilde{\psi}_A s$ by using \eqref{4.8}:
\begin{align}
\psi_B(3)&= \frac{1}{4m^2 c^4} \bigg\{-iD_\tau - m(a\cdot x) - \frac{mc^2}{2}R_{0l0m}x^l x^m +\nonumber 
\\ \qquad &i (\omega \times x)^i D_i - \frac{1}{2} \sigma_p \omega^p \bigg\} \{-\sigma^j ic D_j\} \Tilde{\psi}_A(0) \nonumber\\
 & +\frac{1}{2mc^2} \{-\sigma^j icD_j\} \Tilde{\psi}_A(2) + \frac{1}{12} \{\sigma^b R_{bl0m}x^lx^m \} \Tilde{\psi}_A(1)  \nonumber\\
 &+ \frac{1}{2mc^2} \bigg\{-\sigma^j ic^{-1} (a\cdot x) D_j -\sigma^j \frac{ic}{2}R_{0l0m}x^lx^m D_j - \frac{ic}{6} \sigma^j R_{jl\;\;m}^{\;\;\;i} x^l x^m D_i \nonumber\\
 & - \frac{ic^{-1}}{2}\sigma^j a_j - \frac{ic}{4} \sigma^j R_{0j0l}x^l + \frac{ic}{4} \sigma^j R_{lj}x^l \bigg\} \Tilde{\psi}_A(0) \label{sestar}
\end{align}
Now we use \eqref{4.8}, \eqref{4.14} and \eqref{sestar} to rewrite the equation \eqref{4.16p} at $c^{-2}$ just in terms of $\Tilde{\psi}_A$:
\begin{align}\label{4.21}
  & \left\{ iD_\tau -m(a\cdot x) -\frac{mc^2}{2}R_{0l0m}x^l x^m -i(\omega \times x)^i D_i + \frac{1}{2}\sigma_p \omega^p \right\} \Tilde{\psi}_A(2) \nonumber\\
  & + \left\{\frac{ic}{2}R_{0l\;\;m}^{\;\;\;i}x^l x^m D_i+\frac{ic}{6} \sigma^b \sigma^j R_{bl0m}x^l x^m D_j + \frac{ic}{4} \sigma^b \sigma^j x^l R_{0bjl} \right\} \Tilde{\psi}_A(1) \nonumber\\
  & + \left\{ i \sigma^b \sigma^j x^l (\omega \times x)^r (\frac{1}{8}R_{jblr} + \frac{1}{4}R_{jrbl}) + \frac{i}{24}\omega^p x^l x^m \sigma^b \sigma^j (\epsilon^n_{\;\;pb} R_{jlnm} + \tensor{\epsilon}{^n_{pj}} R_{blnm} ) \right\} \Tilde{\psi}_A(0) \nonumber \\
  &\quad = \frac{1}{4m^2c^4} \{-ic\sigma^j D_j \} \{-iD_\tau\} \{-ic\sigma^j D_j \} \Tilde{\psi}_A(0) \nonumber\\
  &\qquad -\frac{1}{4mc^4} \{-ic\sigma^j D_j \} \{a\cdot x\} \{-ic \sigma^j D_j \} \Tilde{\psi}_A(0) \nonumber \\
  &\qquad - \frac{1}{8mc^2} \{-ic\sigma^j D_j \} \{R_{0l0m}x^l x^m \} \{-ic\sigma^j D_j\}\Tilde{\psi}_A(0) \nonumber\\
  &\qquad+\frac{1}{4m^2c^4}\{-\sigma^j ic D_j\} \{i(\omega \times x)^k D_k\} \{-\sigma^j ic D_j\}\Tilde{\psi}_A(0) \nonumber\\ 
  &\qquad - \frac{1}{8m^2c^4} \sigma^j \sigma_p \sigma^i \omega^p \{ -c^2 D_j D_i \} \Tilde{\psi}_A(0)
  \nonumber\\
  &\qquad- \frac{1}{2m} (\sigma \cdot D)^2 \Tilde{\psi}_A(2) + \frac{1}{12} \{-ic\sigma^j D_j \} \{\sigma^b R_{bl0m}x^lx^m \} \Tilde{\psi}_A(1) \nonumber\\
  &\qquad+  \frac{1}{2mc^2} \{-ic\sigma^j D_j\} \{-ic^{-1} \sigma^j (a\cdot x)D_j \} \Tilde{\psi}_A(0) \nonumber\\
  &\qquad + \frac{1}{2mc^2}\{-ic\sigma^j D_j\} \{-\frac{ic}{2} \sigma^j R_{0l0m}x^l x^m D_j\} \Tilde{\psi}_A(0) \nonumber\\
  &\qquad+  \frac{1}{2mc^2}\{-ic\sigma^j D_j\} \{- \frac{ic}{6} \sigma^j R_{jl\;\;m}^{\;\;\;i}x^l x^m D_i \} \Tilde{\psi}_A(0) \nonumber\\
  &\qquad+ \frac{1}{2mc^2}\{-ic\sigma^j D_j\} \{- \frac{ic^{-1}}{2} \sigma^j a_j\} \Tilde{\psi}_A(0) \nonumber\\
  &\qquad+  \frac{1}{2mc^2}\{-ic\sigma^j D_j\} \{- \frac{ic}{4}\sigma^j R_{0j0l}x^l \} \Tilde{\psi}_A(0) \nonumber\\
  &\qquad+\frac{1}{2mc^2}\{-ic\sigma^j D_j\} \{  \frac{ic}{4}\sigma^j R_{lj}x^l \} \Tilde{\psi}_A(0) \nonumber\\
  &\qquad+ \{- \frac{1}{12} R_{bl0m}\sigma^b x^l x^m \} \{(-ic\sigma^j D_j) \Tilde{\psi}_A(1) +  (\frac{mc^2}{6}\sigma^b R_{bl0m}x^lx^m) \Tilde{\psi}_A(0)\} \nonumber\\
  &\qquad+ \frac{1}{2mc^2} \bigg\{-ic^{-1} \sigma^j (a\cdot x)D_j - \frac{ic}{2} \sigma^j R_{0l0m}x^l x^m D_j - \frac{ic}{6} \sigma^j R_{jl\;\;m}^{\;\;\;i}x^l x^m D_i \nonumber\\
    &\qquad\quad - \frac{ic^{-1}}{2} \sigma^j a_j - \frac{ic}{4} \sigma^j R_{0j0l}x^l + \frac{ic}{4}\sigma^j R_{lj}x^l \bigg\} \{-ic\sigma^j D_j\} \Tilde{\psi}_A(0)
\end{align}
Note that we need to rename some indices in order to continue the calculations. (E.g.\ $\sigma^jD_j$)

Further remark is that the term $\frac{1}{4m^2c^4} \{-ic\sigma^j D_j \} \{-iD_\tau\} \{-ic\sigma^j D_j \} \Tilde{\psi}_A(0)$ which might seem unrelated to the order of approximation, is not actually irrelevant. Because the time derivative of $\Tilde{\psi}_A(0)$ is still of the order of $c^{-2}$ and makes the whole expression again relevant to our approximation. Further details can be found in appendix \ref{appendix E}.

A final remark is in $ \{- \frac{1}{12} R_{bl0m}\sigma^b x^l x^m \} \{(-ic\sigma^j D_j) \Tilde{\psi}_A(1) + (\frac{mc^2}{6}\sigma^b R_{bl0m}x^lx^m) \Tilde{\psi}_A(0)\}$ the second terms is obviously off the order so we will neglect it in the upcoming simplification.
Now we know that in order to read off $H(2)$ we need to look at the operators action on $\Tilde{\psi}_A(0)$ because: 
\begin{align}
i\partial_t \Tilde{\psi}_A(2)=H(2) \Tilde{\psi}_A(0)+ H(1) \Tilde{\psi}_A(1) + H(0) \Tilde{\psi}_A(2)
\end{align}

We can confirm from the above calculations that $H(1)$ and $H(0)$ are calculated correctly and we can read off $H(2)$: 
\begin{align} \label{4.23}
H(2) &= -\frac{1}{4mc^2}a^j D_j - \frac{i}{4mc^2}\tensor{\epsilon}{^{ij}_k}\sigma^k a_i D_j - \frac{1}{4m}R_{0l0m}x^lx^m (\sigma.D)^2 - \frac{1}{2mc^2}(a.x)(\sigma.D)^2 \nonumber\\
&\quad +\frac{1}{8m}R + \frac{1}{4m}R_{00} - \frac{i}{4m}\tensor{\epsilon}{^{ij}_k}\sigma^k R_{0l0i}x^l D_j +\frac{1}{4m}\tensor{R}{_l^j}x^lD_j\nonumber\\
&\quad-\frac{1}{12m}\sigma^i \sigma^j (\tensor{R}{_{jl}^n_i}+\tensor{R}{_{ji}^n_l})x^l D_n - \frac{1}{6m} \delta^i \delta^j R_{jl\;\;m}^{\;\;\;q}x^l x^m D_q D_i  \nonumber\\
  &\quad - \frac{q}{4m^2c^2} \sigma^b \sigma^j D_b E_j   -  \frac{1}{8m^3c^2}(\sigma \cdot D)^4 \nonumber\\
  &\quad + \frac{i}{4m^2c^2}(\omega \times x)^i (\sigma \cdot D)^2 D_i - \frac{i}{4m^2c^2} \sigma^i \sigma^j (\omega \times x)^k D_i D_j D_k \nonumber\\
  &\quad + \sigma^k \tensor{\epsilon}{^{bj}_k}x^l (\omega \times x)^r \Big( \frac{1}{8}\tensor{R}{_{jblr}} +\frac{1}{4} \tensor{R}{_{jrbl}} \Big) - \frac{i}{4} x^l (\omega\times x)^r R_{rl} - \frac{i}{4} x^l (\omega \times x)^r R_{0r0l}
\end{align}
This is the final information that we need in order to calculate the extended Pauli Hamiltonian up to and including the order of $c^{-2}$. Note that in the expression $D_b E_j$ the $D_b$ acts on both the $E_j$ and the $\tilde \psi_A$.
\subsection{Pauli Hamiltonian} 
Now as $H_\text{Pauli} = H(0)+ H(1)+ H(2) + O(c^{-3})$, we conclude (details of the calculation may be found in appendix \ref{appendix E}):
\begin{align} \label{eq4.25}
    H_\text{Pauli} &=  \bigg\{ -\frac{1}{2m} -\frac{1}{2mc^2} (a\cdot x)  - \frac{1}{4m}R_{0l0m}x^l x^m  \bigg\} (D^2 + q(\sigma \cdot B)) \nonumber\\
  &\quad+ \bigg\{- \frac{1}{4mc^2}  a^j - \frac{i}{4mc^2}a_i \tensor{\epsilon}{^{ij}_k}\sigma^k+ \frac{1}{12m}\tensor{R}{_{0l0}^j}x^l - \frac{i}{4m}\tensor{\epsilon}{^{ij}_k}\sigma^k R_{0l0i}x^l \nonumber\\
  &\qquad + \frac{1}{3m} \tensor{R}{_l^j}x^l - \frac{2ic}{3}\tensor{R}{^j_{l0m}}x^lx^m  - \frac{i}{12m} \tensor{\epsilon}{^{iq}_k} \sigma^k x^l (\tensor{R}{_{ql}^j_i} + \tensor{R}{_{qi}^j_l}) + i(\omega \times x)^j \bigg\}D_j \nonumber\\
  &\quad - \frac{1}{6m} \delta^{qi}  \tensor{R}{_{ql}^j_m}x^lx^m D_j D_i \nonumber\\
  &\quad - q c A_0 - \frac{1}{2}\sigma_p \omega^p + \frac{mc^2}{2}R_{0l0m}x^lx^m + \frac{1}{8m}R + \frac{1}{4m}R_{00} + m(a\cdot x)\nonumber\\
  &\quad + \frac{ic}{3}R_{0l}x^l +c \tensor{\epsilon}{^{jb}_k} \sigma^k x^l (\frac{1}{3} R_{0jbl} +\frac{1}{12}R_{bj0l}) \nonumber\\
  &\quad -  \frac{q}{4m^2c^2}((\partial \cdot E) + i\sigma\cdot(\partial\times E)) - \frac{q}{4m^2c^2}\sigma^b \sigma^j E_j D_b  - \frac{1}{8m^3c^2}(\sigma \cdot D)^4 \nonumber\\
  &\quad + \frac{i}{4m^2c^2}(\omega \times x)^i (\sigma \cdot D)^2 D_i -\frac{i}{4m^2c^2} \sigma^i \sigma^j (\omega \times x)^k D_i D_j D_k \nonumber\\
  &\quad + \sigma^k \tensor{\epsilon}{^{bj}_k}x^l (\omega \times x)^r \Big( \frac{1}{8}\tensor{R}{_{jblr}} +\frac{1}{4} \tensor{R}{_{jrbl}} \Big) \nonumber\\
  &\quad- \frac{i}{4} x^l (\omega\times x)^r R_{rl} - \frac{i}{4} x^l (\omega \times x)^r R_{0r0l}+ O(c^{-3})+O(x^{3})
\end{align}
where, we used $\sigma^i \sigma^j = \delta^{ij}\mathbb{1} + i\epsilon^{ij}_{\;\;\;k}\sigma^k$ to simplify most of the terms.
This is the second main result of this thesis.\footnote{It might be asked, why terms related to $c^{-2}m\omega(\omega \times x)$ which stands for centrifugal apparent force does not appear in the Hamiltonian.
In order to answer this, considering the simple case of ordinary classical Hamiltonian mechanics for a point particle, we remember that as we are in the rotating frame, the canonical momentum itself (being expressed in terms of position and velocity) is related to $\omega$. Therefore, we do not have terms related to $\omega^2$ in the Hamiltonian. However, it explicitly appears in the Lagrangian formalism. Therefore the question can be answered by noting the difference between the canonical and kinetic momentum. In this situation they are not equal.
Same argument can be employed for the term related to $c^{-4}(a \cdot x)^2 $.}
This Hamiltonian includes all the terms up to the order of $x^3$, where, $x$ is the distance from the worldline and up to the order of $c^{-3}$, where, $c$ is the speed of light. These are the limits of weak gravity and slow velocities which have been applied separately in two different logically-independent stages of the calculation. We needed \enquote{weak gravity assumption} in order to expand the Dirac equation in the FNC and relate the Dirac equation to the geometry (equation \eqref{eq:Dirac_FNC}). Then we assumed \enquote{slow velocities} in order to study the post-Newtonian limits of the expanded Dirac equation in the FNC. The resulting Hamiltonian is what we called here in the equation \eqref{eq4.25} the Pauli Hamiltonian.

\section{Interpretations}
As we already have the results of the thesis in the equations \eqref{eq:Dirac_FNC} and \eqref{eq4.25}, one might ask; what does this long and somewhat horrible-looking terms refer to? 
To address this question, we need review quickly the steps we have already taken. We first expanded our Dirac equation in Fermi Normal Coordinate which, as we claimed previously, corresponds to our laboratory. Then in order to make sense of Spinors, we used the post-Newtonian paradigm and by the ansatz we chose, we simplified Spinors to wave-functions.

Now the post-Newtonian effects and corrections which appear in the so-to-say Pauli Hamiltonian at hand \eqref{eq4.25} are ready to be interpreted. 
Interpretation is important here, because it will guide experimental physicists to look for the right effects and attempt to observe those. Only after the correct interpretation, experimental observation can confirm whether the approach we have used is correct. 

The interpretation of terms as they appear in one of the resulting Dirac/Pauli Hamiltonian is still an ongoing research and is beyond the scope of this thesis. However, we can take two simple examples and see how the process works.  

As the first example, in order to translate back the mathematics to its realistic correspondence, we recall that $\sigma^i$ corresponds to Spins. Therefore terms with $\sigma^i$ are the spin coupling terms. For example, $-\frac{1}{2}\sigma_p \omega^p$ is the coupling of the spin to the angular velocity of the system which modifies the magnetic moment i.e.\ the $g$-factor \cite{PhysRevLett.61.2639}. 

As the second example, considering the coupling of the Newtonian gravitational potential to the Schrödinger equation:
\begin{align}
    i\partial_t \psi = \left(\frac{p^2}{2m}+m\phi\right)\psi
\end{align}

we can conclude that Newtonian gravitational potential can be expressed as the coefficient of $m$ in the Pauli Hamiltonian. Therefore we can name the following terms as the Newtonian gravitational potential:
\begin{align}
   \phi= c^{-2}(a\cdot x) +\frac{1}{2}R_{0l0m}x^l x^m +O(x^3) + O(c^{-3})
\end{align}
Besides, knowing the Hessian as following:
\begin{align}
    \phi(x_0 + x)=\phi(x_0)+x\cdot\nabla\phi(x_0)+\frac{1}{2}x^i x^j (\partial_i \partial_j \phi)(x_0)+ O(x^3)
\end{align}
where, 
\begin{align}
     (\partial_i \partial_j \phi)(x_0) = ((H\phi)_{(x_0)})_{ij}
\end{align}
we can see the term $R_{0l0m}$ in the Hamiltonian is the Hessian.

Apart from these examples, we could have also translated the terms in terms of physical operators as $H=H({x,P,S})$, where $S$ is the spin operator, $P$ is the momentum operator and $x$ is the position operator; but we will leave it for section \ref{okas} where we discuss this approach in detail.

Finally, I need to mention that, in March 1978, Ni and Zimmermann published a paper titled as \enquote{Inertial and gravitational effects in the proper reference frame of an accelerated, rotating observer} \cite{q:PhysRevD.17.1473}. There, the authors expanded the equation of motion of a freely falling particle in the FNC and then they tried to interpret the terms. For the further interpretation one could refer to this paper.

\chapter{Comparison to previous works} \label{chap5}

In this chapter, I will compare my results with two papers which have been published recently \cite{perche21,Ito:2020xvp}. The comparisons are formulated in the context of two independent `reports', which are the sections of this chapter.  \footnote{This chapter is revised and updated after fruitful scientific debates with the authors of both papers.}

\section[Perche \& Neuser 2021]{Perche \& Neuser 2021: \enquote{A wavefunction description for a localised quantum particle in curved spacetimes}}

In a paper titled \enquote{A wavefunction description for a localised quantum particle in curved spacetimes} \cite{perche21}, Perche \& Neuser give a well-thought out description of the Dirac Hamiltonian on expanded metric in Fermi Normal Coordinates. Here, I will try to point out and clarify some of the most important differences of their method and ours.

We will show that our method of post-Newtonian expansion is more general than what they applied. Our method leads to extra terms in the post-Newtonian resulting Hamiltonians (Pauli and Schrödinger Hamiltonians) which are of the order of $m^{-1}$ and neglected due to the assumptions of their method. Finally, we will bring an argument named \enquote{Tracing or not?} in order to see whether the intention behind the reduction of post-Newtonian Pauli Hamiltonian to the Schrödinger Hamiltonian is justifiable. 

\subsection{Comparing the Dirac Hamiltonian }

I calculated the Dirac Hamiltonian for a general metric which is expanded in a rotating Fermi Normal Coordinates with an arbitrary acceleration up to the order of $x^2$. In order to make my calculation comparable with Perche-Neuser, I set $\omega$ (the angular velocity, which causes the rotation in the FNC) to zero, and I ignore all the non-linear terms of $a$ and $R_{\mu\nu\rho\sigma}$. The result is as follows:
\begin{align}
 H_{\text{Dirac}}=&\gamma^0 \bigg\{mc^2+m(a\cdot x) +\frac{mc^2}{2}R_\mathrm{0l0m}x^l x^m\bigg\}\nonumber \\
& -\gamma^\mathrm{b} \bigg\{\frac{mc^2}{6}R_\mathrm{bl0m}x^lx^m\bigg\}- \mathbb{1}\bigg\{\frac{ic}{2}R^i_{\mathrm{\;m0l}} x^lx^m\partial_i \bigg\} \nonumber \\
&-\gamma^\mathrm{0} \gamma^\mathrm{j} \bigg\{\delta^i _\mathrm{j}(ic\partial _i+ic^{-1}(a\cdot x)\partial _i+\frac{ic}{2}R_\mathrm{0l0m}x^lx^m\partial_i) +\frac{ic}{6}R^i_\mathrm{\;mjl}x^lx^m\partial_i+\frac {ic^{-1}}{2}a_\mathrm {j} \nonumber \\
&+\frac {ic}{4} R_\mathrm{0j0l}x^l-\frac {ic}{4}R_{\mathrm{lj}}x^l \bigg\}+\gamma^\mathrm{b} \gamma^\mathrm{j} \bigg\{\delta ^i _{\mathrm{j}}(\frac {ic}{6}R_{\mathrm{bl0m}}x^l x^m \partial _i ) +\frac{ic}{4}x^l R_{\mathrm{objl}}\bigg\}
\end{align}
It is identical to the equation $62$ in Perche \& Neuser's paper \cite{perche21}. However, the difference between the post-Newtonian expansion methods will later cause a rather important difference in the post-Newtonian Pauli Hamiltonian.

\subsection{Differences of $1/m$ and $1/c$ expansions}
In order to see the post-Newtonian limit\footnote{It is not exactly what they intended to do. However, one can use it as their calculations' side result.}, Perche $\&$ Neuser split the Dirac Spinor into positive and negative $\psi$s as follows:
\begin{align}
  \psi= \begin{pmatrix} \psi_A \\ \psi_B \end{pmatrix} \\
\end{align}
Then they can write down the Hamiltonian as:
\begin{align}
  H= \begin{pmatrix} H_{AA} \;\; H_{AB}\\ H_{BA}\;\; H_{BB} \end{pmatrix}
\end{align}
Therefore the Dirac equation can be broken into the following equations:
\begin{align}
&i\partial_t\psi_A=H_{AA}\psi_A + H_{AB}\psi_B \\
&i\partial_t\psi_B=H_{BA}\psi_A + H_{BB}\psi_B
\end{align}
One can read off each $H$ from the given Dirac Hamiltonian. 
Then they can write down the second equation as follows:
\begin{align}
&\psi_B = D_B \psi_A \\
&D_B:= (i\partial_t- H_{BB})^{-1} H_{BA}
\end{align}
Therefore the equations for the states are decoupled:
\begin{align}
    id_t \psi_A = (H_{AA}+H_{AB}D_B)\psi_A
\end{align}
$(H_{AA}+H_{AB}D_B)$ will be then our post-Newtonian Pauli $H$ including spins, if one calculates the part $(id_\tau- H_{BB})^{-1}$ in $ D_B$ by defining $id_T=i\partial_\tau - mc^2$  and uses geometric series and expands it in $1/m$. Where, $id_T$ is so called \enquote{energy operator}. 
Referring to their paper: 

In equation C2 to C3 they apply what was explained above. And in transition of equation C3 to C4 they simply argue that they keep the order of approximation up to and including $1/m$ in $(i\partial_\tau- H_{BB})^{-1}$, which is \emph{just} part of $ D_B$ .
That is why the term $ \frac{id_T+q c A_0}{2mc^2}$ is eliminated. However, neglecting this term will lead to differences at order $m^{-1}$ in the end. More explicitly, $D_B$ must have had the term $ \frac{id_T+q c A_0}{4m^2c^4}(-ic(\sigma.D))$ in order not to miss any information of the order of $1/m$ and to be mathematically consistent.\footnote{The mentioned term might seem off the order of approximation but as we continue it turns out that it is relevant. It is to be explained in the followings.}Then $D_B$ would then be: 
\begin{align}\label{pnm}
D_B &= \frac{1}{2mc^2} \bigg\{1-\frac{id_T + q c A_0}{2mc^2} -\frac{c^{-2}}{2}(a\cdot x)- \frac{1}{4}R_{0l0m}x^lx^m \bigg\}H_{BA} \nonumber \\ 
&= \frac{1}{2mc^2} \bigg\{1 -\frac{c^{-2}}{2}(a\cdot x)- \frac{1}{4}R_{0l0m}x^lx^m \bigg\}H_{BA}  + O
\end{align}
where, $O$ is:
\begin{align}
    O&=-\frac{id_T+q c A_0}{4m^2c^4} H_{BA} \nonumber\\
    &=-\frac{id_T+q c A_0}{4m^2c^4}\bigg\{ \frac{mc^2}{6}\sigma^b R_{bl0m}x^lx^m-ic(\sigma.D) -ic^{-1}(a.x)(\sigma.D)-\frac{ic}{2}R_{0l0m}x^lx^m (\sigma.D)-\nonumber\\
    &\qquad \qquad \qquad \frac{ic}{6} \sigma^j R_{jl\;m}^{\;\;\;i}x^l x^m D_i - \frac{ic^{-1}}{2}\sigma^j a_j - \frac{ic}{4}R_{0j0l}x^l \sigma^j + \frac{ic}{4}\sigma^j R_{lj}x^l \bigg\}
\end{align}
Therefore:
\begin{align}
&i\partial_t \psi_A = (H_{AA}+H_{AB}D_B)\psi_A \\ \label{pnm3}
&\implies i\partial_t \psi_A \sim H_{AB}O
\end{align}
Now, let us see what is the explicit form of $H_{AB}O$:
\begin{align}
    H_{AB}O&= -\bigg\{- \frac{mc^2}{6}\sigma^b R_{bl0m}x^lx^m - \sigma^j ic D_j - ic^{-1}(a\cdot x)\sigma^j D_j - \frac{ic}{2} R_{0l0m}x^l x^m \sigma^j D_j \nonumber \\
&- \frac{ic}{6} \sigma^j R_{jl\;m}^{\;\;\;i}x^l x^m D_i - \frac{ic^{-1}}{2}\sigma^j a_j - \frac{ic}{4}R_{0j0l}x^l \sigma^j + \frac{ic}{4}\sigma^j R_{lj}x^l \bigg\}O \nonumber\\
    &=\bigg\{- \frac{mc^2}{6}\sigma^b R_{bl0m}x^lx^m - \sigma^j ic D_j - ic^{-1}(a\cdot x)\sigma^j D_j - \frac{ic}{2} R_{0l0m}x^l x^m \sigma^j D_j \nonumber \\
&- \frac{ic}{6} \sigma^j R_{jl\;m}^{\;\;\;i}x^l x^m D_i - \frac{ic^{-1}}{2}\sigma^j a_j - \frac{ic}{4}R_{0j0l}x^l \sigma^j + \frac{ic}{4}\sigma^j R_{lj}x^l \bigg\}\frac{id_T+q c A_0}{4m^2c^4}\nonumber\\
&\bigg\{ -ic(\sigma.D)-ic^{-1}(a.x)(\sigma.D)-\frac{ic}{2}R_{0l0m}x^lx^m (\sigma.D)-\frac{ic}{6} \sigma^j R_{jl\;m}^{\;\;\;i}x^l x^m D_i -\nonumber\\ & \frac{ic^{-1}}{2}\sigma^j a_j -   \frac{ic}{4}R_{0j0l}x^l \sigma^j + \frac{ic}{4}\sigma^j R_{lj}x^l \bigg\}
\end{align}

Note that as $id_T=id_\tau -m$, the other terms (\emph{which can also be of the order of $\frac{1}{m^2}$}) in \eqref{pnm} also contribute to the calculation. Having the order of approximation $\frac{1}{m}$, the neglected term in their calculation is:
\begin{align} \label{wwwo}
    H_{AB}O&= -\{ic\sigma.D\}(\frac{iD_T}{4m^2c^4})\{ic\sigma.D\}
\end{align}
where, $iD_T=id_T+q c A_0$. We can now rewrite the equation \eqref{wwwo} using $iD_T=iD_\tau-mc^2$:
\begin{align} \label{wwwoo}
    H_{AB}O&= \frac{1}{4m^2c^2}\{\sigma.D\}(iD_\tau)\{\sigma.D\} -\frac{1}{4m}(\sigma.D)^2 \nonumber\\
    \implies H_{AB}O&=-\frac{1}{8m^2c^2}(\sigma.D)^2(iD_\tau)+ \frac{1}{8m^2c^2}(\sigma.D)[(\sigma.D),iD_\tau]+\frac{1}{8m}(\sigma.D)^2 \nonumber\\
    \implies H_{AB}O&= -\frac{1}{8m^2c^2}(\sigma.D)^2 H_{\text{Pauli}}+ \frac{1}{8m^2c^2}(\sigma.D)[(\sigma.D),iD_\tau]+\frac{1}{8m}(\sigma.D)^2
\end{align}
where, $[,]$ is the commutation relation. In the last step we replaced the first term using the fact that $iD_\tau \psi= H\psi$. Note that the second term is off the order of approximation. To write the neglected terms down explicitly, we can label them as $E$:
\begin{align} \label{E}
E:= \frac{1}{2mc^2}a^jD_j + \frac{1}{4mc^2}(a.x)(\sigma.D)^2 + \frac{1}{8m}R_{0l0m}x^lx^m (\sigma.D)^2 +\frac{1}{4m}R_{00}+\frac{1}{2m}\tensor{R}{_0^j_{0m}}x^m D_j
\end{align}

Note that this is exactly what I have calculated in appendix \ref{appendix E}. These new terms are obtained if one does not assume that the system's non-relativistic energy is negligible compared to its rest mass. This is where we claim that our post-Newtonian method is more general than their method, as it requires less assumptions and produces more terms.

\subsection{Perche \& Neuser's corrected Pauli and Schrödinger Hamiltonian}\footnote{The main reason for including this section in my thesis and not just referring to their paper is that they do not calculate the Pauli Hamiltonian explicitly, and do not provide full details of their calculation.}
Here, I calculate the corrected Hamiltonian doing the calculation with the $1/m$ expansions.
For doing so, we will first write down $H_{AA}$ , $H_{AB}$ , $H_{BA}$ and $H_{BB}$ explicitly: 
\begin{align}
H_{AA}=&- \frac{ic}{2}R_{0l\;m}^{\;\;\;i\;}x^l x^m D_i + mc^2 +m(a\cdot x) + \frac{mc^2}{2} R_{0l0m}x^l x^m  \nonumber \\
&-\sigma^b \sigma^j \frac{ic}{6}R_{bl0m}x^l x^m D_j -
\sigma^b \sigma^j \frac{ic}{4} x^l R_{0bjl}
\\[1em]
H_{AB}=& - \frac{mc^2}{6}\sigma^b R_{bl0m}x^lx^m - \sigma^j ic D_j - ic^{-1}(a\cdot x)\sigma^j D_j - \frac{ic}{2} R_{0l0m}x^l x^m \sigma^j D_j \nonumber \\
&- \frac{ic}{6} \sigma^j R_{jl\;m}^{\;\;\;i}x^l x^m D_i - \frac{ic^{-1}}{2}\sigma^j a_j - \frac{ic}{4}R_{0j0l}x^l \sigma^j + \frac{ic}{4}\sigma^j R_{lj}x^l 
\\[1em]
H_{BA}=&  \frac{mc^2}{6}\sigma^b R_{bl0m}x^lx^m - \sigma^j ic D_j - ic^{-1}(a\cdot x)\sigma^j D_j - \frac{ic}{2} R_{0l0m}x^l x^m \sigma^j D_j \nonumber\\
&- \frac{ic}{6} \sigma^j R_{jl\;m}^{\;\;\;i}x^l x^m D_i - \frac{ic^{-1}}{2}\sigma^j a_j - \frac{ic}{4}R_{0j0l}x^l \sigma^j + \frac{ic}{4}\sigma^j R_{lj}x^l 
\\[1em]
H_{BB}=&- \frac{ic}{2}R_{0l\;m}^{\;\;\;i\;}x^l x^m D_i - mc^2 -m(a\cdot x) - \frac{mc^2}{2} R_{0l0m}x^l x^m \nonumber\\
&- \sigma^b \sigma^j \frac{ic}{6}R_{bl0m}x^l x^m D_j - \sigma^b \sigma^j \frac{ic}{4} x^l R_{0bjl}
\end{align}
Therefore one can calculate $D_B$: 
    \begin{align}
D_B &= \frac{1}{12} \sigma^b R_{bl0m}x^lx^m - \frac{i}{2mc} \sigma^j \left(1+ \frac{c^{-2}}{2}(a\cdot x) + \frac{1}{4} R_{0l0m}x^l x^m \right) D_j \nonumber\\
&\quad - \frac{1}{2mc^2} \left(\frac{ic}{6}R_{jl\;m}^{\;\;\;i}x^l x^m \sigma^j D_i + \frac{ic^{-1}}{2} \sigma^j a_j + \frac{ic}{4}\sigma^j R_{0j0l}x^l - \frac{ic}{4}\sigma^j R_{lj}x^l \right)
    \end{align}
Finally, we need $H_{AB}D_B$: 
    \begin{align} \label{pmpm}
H_{AB}D_B &= \frac{ic}{12}[ \sigma^b , \sigma^j ] R_{bl0m}x^l x^m D_j - \frac{ic}{12} \sigma^j \sigma^b (R_{bj0l}+ R_{bl0j})x^l \nonumber\\
&\quad - \frac{1}{2m} \left(1+\frac{3c^{-2}}{2}(a\cdot x)+ \frac{3}{4}R_{0l0m}x^l x^m \right) (D.\sigma)^2 \nonumber\\
&\quad - \frac{1}{2m} \sigma^j \sigma^q \left(\frac{c^{-2}}{2}a_j + \frac{1}{2}R_{0j0l}x^l \right) D_q - \frac{1}{12m}\sigma^j \sigma^q (R_{qj\;l}^{\;\;\;i}+ R_{ql\;j}^{\;\;\;i})x^l D_i \nonumber\\
&\quad -\frac{1}{8m}\sigma^j \sigma^q R_{0q0j} + \frac{1}{8m} R_{lq}x^l \{\sigma^j ,\sigma^q\}D_j + \frac{1}{8m} \sigma^j \sigma^q R_{jq} \nonumber\\
&\quad - \frac{1}{12m} R_{jl\;m}^{\;\;\;i}x^l x^m \{ \sigma^j D_i , \sigma^q D_q \} - \frac{1}{4mc^2} a_q \{ \sigma^j , \sigma^q \}D_j - \frac{1}{8m} R_{0j0l} \{ \sigma^j , \sigma^q \} D_q
    \end{align}
where, we used square brackets and curly brackets to denote commutators and anti-commutators, respectively, in order to simplify the equation.\footnote{
A very important note is that in equation \eqref{pmpm}, it is one of the rare places we use such notation. Later to make the calculations more understandable we will use \enquote{$\{$} and \enquote{$\}$} as brackets.}
Now we name $H_{AA}+H_{AB}D_B$ as the post-Newtonian Pauli Hamiltonian. 
    \begin{align} \label{62444}
H_\text{Pauli}+E= H_\text{Pauli by me in their level of approx.}
\end{align}
where, $E$ is calculated in \eqref{E}.

\subsection{Comparing the Pauli Hamiltonians}
My Pauli Hamiltonian in the order of approximation $m^{-1}$ and linear in $R$ and $a$, while $\omega$ is vanishing, is: 
    \begin{align} \label{5.21}
    H_\text{Pauli} &=  \bigg\{ -\frac{1}{2m} -\frac{1}{2mc^2} (a\cdot x)  - \frac{1}{4m}R_{0l0m}x^l x^m  \bigg\} (D^2 + q (\sigma \cdot B)) \nonumber\\
  &\quad+ \bigg\{- \frac{1}{4mc^2}  a^j - \frac{i}{4mc^2}a_i \tensor{\epsilon}{^{ij}_k}\sigma^k+ \frac{1}{12m}\tensor{R}{_{0l0}^j}x^l - \frac{i}{4m}\tensor{\epsilon}{^{ij}_k}\sigma^k R_{0l0i}x^l \nonumber\\
  &\qquad + \frac{1}{3m} \tensor{R}{_l^j}x^l - \frac{2ic}{3}\tensor{R}{^j_{l0m}}x^lx^m  - \frac{i}{12m} \tensor{\epsilon}{^{iq}_k} \sigma^k x^l (\tensor{R}{_{ql}^j_i} + \tensor{R}{_{qi}^j_l}) \bigg\}D_j \nonumber\\
  &\quad - \frac{1}{6m} \delta^{qi}  \tensor{R}{_{ql}^j_m}x^lx^m D_j D_i \nonumber\\
  &\quad -q c A_0  + \frac{mc^2}{2}R_{0l0m}x^lx^m + \frac{1}{8m}R + \frac{1}{4m}R_{00} + m(a\cdot x)\nonumber\\
  &\quad + \frac{ic}{3}R_{0l}x^l +c \tensor{\epsilon}{^{jb}_k} \sigma^k x^l \left(\frac{1}{3} R_{0jbl} +\frac{1}{12}R_{bj0l}\right)
\end{align}
    
Comparing it to what Perche \& Neuser calculated in their paper \cite{perche21}, we note that the terms labelled as $E$ in equation \eqref{E} are neglected in their Pauli Hamiltonian.

\subsection{Comparing the Schrödinger Hamiltonians}
Now we can, finally, trace over spin degrees of freedom (using the Pauli matrices algebra as exactly what Perche \& Neuser did) in order to calculate the Schrödinger Hamiltonian as Perche-Neuser did. The result then would be Post-Newtonian Schrödinger Hamiltonian in curved spacetime:
\begin{align}\label{Schrödinger 2}
H_\text{Schrödinger}&=  \bigg\{ -\frac{1}{2m} -\frac{1}{2mc^2} (a\cdot x)  - \frac{1}{4m}R_{0l0m}x^l x^m  \bigg\} D^2 - \frac{1}{6m}  \tensor{R}{^i_{l}^j_m}x^lx^m D_j D_i \nonumber\\
  &\quad- \bigg\{ \frac{1}{4mc^2}  a^j - \frac{1}{12m}\tensor{R}{_{0l0}^j}x^l - \frac{1}{3m} \tensor{R}{_l^j}x^l + \frac{2ic}{3}\tensor{R}{^j_{l0m}}x^lx^m \bigg\}D_j \nonumber\\
  &\quad   -q c A_0+ \frac{mc^2}{2}R_{0l0m}x^lx^m + \frac{1}{8m}R + \frac{1}{4m}R_{00} + m(a\cdot x) + \frac{ic}{3}R_{0l}x^l
\end{align}

Comparing it to Perche \& Neuser's Schrödinger Hamiltonian, we note that the terms labelled as $E$ \eqref{E} are neglected in their Schrödinger Hamiltonian.

Now there are few remarks about what we have achieved. First, Dirac Hamiltonians were identical. As we went further, we realised that new terms arising from my different assumptions start having more and more consequences. Therefore we decided to explicitly calculate the term they neglected. We concluded at the end that all the differences between our calculation and theirs are caused by the less general choice of post-Newtonian approximation method they are applying. 

In addition, tracing over spin degrees of freedom is still ambiguous as we just trace the terms which include sigma terms and we leave out the terms with the identity matrix in front of them. We will discuss it in more details in the following part.

\subsection{Tracing or not?}

As we saw in the previous sections, in order to calculate the effective Schrödinger Hamiltonian from the post-Newtonian Pauli Hamiltonian, Perche \& Neuser traced over the spin degrees of freedom, such that only the terms corresponding to the sigma matrices were affected. This method is, however, not physically meaningful: the unitary time evolution described by the full post-Newtonian Pauli Hamiltonian will induce interactions between the position and spin degrees of freedom. Therefore, the time evolution which we would obtain by ignoring the spin, i.e.\ by taking the partial trace of the total density matrix over the spin degrees of freedom, would no longer be unitary. Therefore, it cannot be described by any Schrödinger equation with respect to a Hamiltonian. Of course, it might be possible to argue for an approximately unitary time evolution in some cases, but Perche \& Neuser don't do this.

\subsection{Summary}
To summarise, in this section we contend that Perche \& Neuser's method to reduce the Dirac Hamiltonian to the Pauli Hamiltonian is not complete if understood as a $\frac{1}{m}$ expansion, because a term relevant at order $\frac{1}{m}$ was neglected by them: we found out that if one wants to consistently expand in $\frac{1}{m}$, there are crucial terms missing in their post-Newtonian Pauli Hamiltonian, such as $\frac{1}{4m}R_{00}$. The extra terms arise if one does not assume that the non-relativistic energy of the system is negligible compared to its rest mass, which Perche \& Neuser did. This is a novel result of our calculation, when compared to previous literature, but also expected, due to the fewer assumptions we make in our post-Newtonian method. These terms could have been reproduced by applying their method and not neglecting the term in \eqref{E}, thus turning their calculation into a systematic expansion in $\frac{1}{m}$ (instead of an expansion neglecting terms assumed to be small compared to rest mass).\footnote{One could argue that Perche \& Neuser did not consider the $\frac{1}{4m} R_{00}$ term, because they assumed that their calculation is outside of the matter, i.e.\ they assumed to be in the situation where $R_{\mu\nu}$-related terms are vanishing. This could have been a valid argument, if they did not have the Ricci term neither. However, they do have the term related to $\frac{1}{m}R$ in their Hamiltonians and this makes the aforementioned argument invalid. }
Furthermore, the idea of reducing the Pauli Hamiltonian to the Schrödinger Hamiltonian by tracing over the spin degrees of freedom seems also conceptually unsound.

\section[Ito 2021]{Ito 2021: \enquote{Inertial and gravitational effects\\ on a geonium atom}} \label{okas}

This section is to compare my result with what Ito calculated in a paper titled \enquote{Inertial and gravitational effects on a geonium atom} \cite{Ito:2020xvp}, where he uses the FW-transformation \cite{PhysRev.78.29} in order to study the post-Newtonian limit of Dirac Hamiltonian.
Again, we will call such a limit, \enquote{Pauli Hamiltonian} and we will check whether the results are comparable. We will see that one needs to apply a Lorentz transformation due to the different choice of Fermi Normal basis vectors. We will interpret the geometric meaning of the possible choices and conclude that his choice might not be the best.  
After finding the Lorentz transformation, in the next section, we will transform our result to a Hamiltonian which would be comparable to Ito's Pauli Hamiltonian. We will see that the results are still incomparable. We will discuss the possible reasons for such differences and conclude the section by stating that due to the different methods of post-Newtonian approximation, the resulting Pauli Hamiltonians seem to be impossible to be compared.

\subsection{Comparing the Dirac Hamiltonian}

Ito calculated the Dirac Hamiltonian to the linear order of curvature, acceleration and angular velocity. Converting his notation to mine and including all the c-factors, we can see that the resulting Hamiltonian in his paper is:
\begin{align}
    H_{\text{Dirac}} =& \mathbb{1} \bigg\{-q c A_0-\frac{ic}{2}\tensor{R}{_{0l}^i_m}x^lx^mD_i \bigg\} + \gamma^0 \bigg\{mc^2 +m(a\cdot x) +\frac{mc^2}{2}R_{0l0m}x^lx^m \bigg\}\nonumber\\
    &- \gamma^b \bigg\{ mc(\omega \times x)_b + \frac{mc^2}{6}R_{bl0m}x^lx^m\bigg\} \nonumber\\
    &-\gamma^0 \gamma^{\mathrm{j}}\bigg\{ icD_j + ic^{-1}(a\cdot x)D_j +\frac{ic}{2}R_{0l0m}x^lx^mD_j +\frac{ic}{6}\tensor{R}{_{jl}^i_m}x^lx^m D_i + \frac{ic^{-1}}{2}a_j \nonumber\\
    &+\frac{ic}{4}R_{0j0l}x^l - \frac{ic}{4}R_{lj}x^l\bigg\} +\gamma^b\gamma^{\mathrm{j}}\bigg\{ \frac{ic}{6}R_{bl0m}x^lx^m D_j +\frac{ic}{4}R_{0bjl}x^l + i(\omega \times x)_b D_j\bigg\}
\end{align}
It is worth seeing my result in the same order of approximation and compare the two Dirac Hamiltonians. My Hamiltonian is:
\begin{align}
    H_{\text{Dirac}} =& \mathbb{1} \bigg\{i(\omega \times x)^iD_i-q c A_0-\frac{ic}{2}\tensor{R}{_{0l}^i_m}x^lx^mD_i \bigg\} \nonumber\\
    &+ \gamma^0 \bigg\{mc^2 +m(a\cdot x) +\frac{mc^2}{2}R_{0l0m}x^lx^m \bigg\}- \gamma^b \bigg\{\frac{mc^2}{6}R_{bl0m}x^lx^m\bigg\} \nonumber\\
    &-\gamma^0 \gamma^{\mathrm{j}}\bigg\{ icD_j + ic^{-1}(a\cdot x)D_j +\frac{ic}{2}R_{0l0m}x^lx^mD_j +\frac{ic}{6}\tensor{R}{_{jl}^i_m}x^lx^m D_i + \frac{ic^{-1}}{2}a_j \nonumber\\
    &+\frac{ic}{4}R_{0j0l}x^l - \frac{ic}{4}R_{lj}x^l\bigg\} +\gamma^b\gamma^{\mathrm{j}}\bigg\{ \frac{ic}{6}R_{bl0m}x^lx^m D_j +\frac{ic}{4}R_{0bjl}x^l + \frac{i}{4}\epsilon_{bpj} \omega^p \bigg\}
\end{align}
As we can see, there are just two terms which are different and they are both related to angular velocity. More explicitly, in Ito's Hamiltonian, they are the first term in $\gamma^b$ and last term in $\gamma^b \gamma^{\mathrm{j}}$ which take different forms than my Hamiltonian's first term in $\mathbb{1}$ and last term in $\gamma^b \gamma^{\mathrm{j}}$.
The reason is the different choice of basis vectors for expanded metric in Fermi Normal Coordinate.\footnote{If one considers $\omega=0$ then the resulting Dirac Hamiltonian are identical here, this is a good reason to be confident about our result and the comparison we did in the previous section.  }

\subsubsection{Different choice of basis vectors}

Choosing the basis vectors in the first glance, would seem to be unimportant. It is the matter of the taste and simplicity which may be brought for the rest of calculation. However, when it comes to geometric description and more importantly, the post-Newtonian approximations, it is really important to choose the basis vectors carefully. 

Ito chooses his basis vectors as follows:
\begin{align}
 &e^0_s= 1+c^{-2}(a\cdot x) +\frac{1}{2}R_{0k0l}x^kx^l \nonumber\\ 
&e^{\mathrm{i}}_s=- \frac{1}{2}R^i_{\;k0l}x^k x^l \nonumber\\ 
&e^0_j=c^{-1} \epsilon_{jmk} x^m \omega^k + \frac{1}{6}R_{0kjl}x^k x^l \nonumber\\ 
&e^{\mathrm{i}}_j =\delta^i_j - \frac{1}{6}R^i_{\;kjl}x^k x^l 
\end{align}
He did not give any argument for justifying his choice.
On the other hand, my choice of basis vector was as follows:
\begin{align}
    &e^0_s=1+c^{-2}(a\cdot x)+\frac{1}{2}R_{0l0m}x^lx^m \nonumber \\
    &e^{\mathrm{i}}_s= - \frac{1}{2}\tensor{R}{^{\mathrm{i}}_{l0m}}x^lx^m +c^{-1}(\omega \times x)^{\mathrm{i}} \nonumber \\
    &e^0_j= \frac{1}{6}R_{0ljm}x^lx^m \nonumber \\
    &e^\mathrm{i}_j= \delta^{\mathrm{i}}_j -\frac{1}{6}\tensor{R}{^{\mathrm{i}}_{ljm}}x^lx^m 
\end{align}
As it can be seen, up to the choice of angular velocity being a part of $e^0_j$ or $e^{\mathrm{i}}_s$ they are identical.
In order to give an argument we also need to calculate the dual basis;
Ito's dual basis are:
\begin{align}
    &e^s_0=1-c^{-2}(a\cdot x)-\frac{1}{2}R_{0l0m}x^lx^m \nonumber\\
    &e^s_{\mathrm{i}}=-\frac{1}{6}\tensor{R}{_{il0m}}x^lx^m + c^{-1}(\omega \times x)_i \nonumber\\
    &e^j_0= \frac{1}{2} \tensor{R}{_{0l}^j_m}x^lx^m \nonumber\\
    &e^j_{\mathrm{i}}=\delta^j_{\mathrm{i}}+\frac{1}{6}\tensor{R}{_{il}^j_m}x^lx^m
\end{align}
My dual basis up to same level of approximation as Ito are:
\begin{align}
    &e^s_0=1-c^{-2}(a\cdot x)-\frac{1}{2}R_{0l0m}x^lx^m \nonumber \\
    &e^s_{\mathrm{i}}= - \frac{1}{6}\tensor{R}{_{\mathrm{i}}_{l0m}}x^lx^m  \nonumber \\
    &e^j_0= -c^{-1}(\omega \times x)^j  +\frac{1}{2}\tensor{R}{_{0l}^j_m}x^lx^m \nonumber \\
    &e^j_{\mathrm{i}}= \delta^j_{\mathrm{i}} +\frac{1}{6}\tensor{R}{_{\mathrm{i}}_l^j_m}x^lx^m 
\end{align}

\begin{figure}
\begin{center}
\begin{tikzpicture}
\draw (0,0).. controls (0,.5) and (.5,0) .. (1,1);\draw   (1,1) ..controls (2,.5) and (2,1).. (4,1) ; \draw (4,1)..controls (3,.5)..node[right] {\quad$\tau=cte$}(2.5,0); \draw (2.5,0)..controls (1,-.5) and (.1,.3)..(0,0);

\draw[very thick,-stealth] (1.5,.5) -- node[right] {$\vec v$} (1.6,1.5);
\draw (1.5,.68)--(1.7,.70); \draw (1.7,.7)--(1.68,.5);
\end{tikzpicture}
\end{center}
\caption{Choosing $\vec v$  orthogonal to the spatial leaf of the constant $\tau$}
\label{fig:1}
\end{figure}
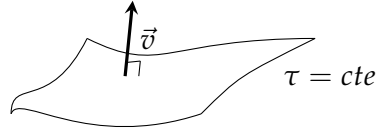

Having calculated the basis and their dual, one can see whether they correspond to something meaningful in geometry. This is done by looking at the geometric definition of time in the neighbourhood of the worldline. We know that there are two possible choices for defining  the measurable usual time which is corresponding to the clock in a laboratory in the neighbourhood of the worldline. 
The first option is to define the time to be the normal vector which is orthogonal to the hyper spatial plane of $\tau=constant$, where $\tau$ is the proper time (see figure \ref{fig:1}). In this case, mathematically we have:
\begin{align}
    (grad(\tau))^\mu =g^{\mu \nu}(d\tau)_\nu
\end{align}
After normalising it, we have:
\begin{align}
    e^\mu_0=- \frac{g^{\mu s}}{(-g^{ss})^{\frac{1}{2}}}
\end{align}
Note that the minus sign is there to make the vector future directed.

\begin{figure}
\begin{center}
\begin{tikzpicture}
\draw (0,0).. controls (0,.5) and (.5,0) .. (1,1);\draw   (1,1) ..controls (2,.5) and (2,1).. (4,1) ; \draw (4,1)..controls (3,.5)..node[right] {\quad$\tau=\tau_1$}(2.5,0); \draw (2.5,0)..controls (1,-.5) and (.1,.3)..(0,0);

\draw (0,-2).. controls (0,-1.5) and (.5,-2) .. (1,-1);\draw   (1,-1) ..controls (2,-1.5) and (2,-1).. (4,-1) ; \draw (4,-1)..controls (3,-1.5)..node[right] {\quad$\tau=\tau_0$}(2.5,-2); \draw (2.5,-2)..controls (1.3,-1.7) and (.1,-1.7)..(0,-2);

\filldraw [black] (2,.5) circle (1pt);
\draw[gray] (2,.5)--node[left] {$(\tau_1,x^i)$\qquad\;\;\;\;\quad\;}(-1,.5);

\filldraw[black]  (2,-1.5) circle (1pt);
\draw[gray] (2,-1.5)--node[left] {$(\tau_0,x^i)$\qquad\;\;\;\;\quad\;}(-1,-1.5);

\draw[thick] (2.2,0)..controls (2.4,-.7) and (1.8,-.3)..(2,-1.5);
\draw[thick,dotted] (2.2,0)--(2,.5);
\draw[thick] (2,.5)--(1.7,1.2);
\draw[thick,dotted] (2,-1.5)--(2.1,-1.9);
\draw [thick] (2.1,-1.9)--(2.3,-2.4);

\draw [very thick,-stealth] (2,-1.5)--node[left]  {\;$\Vec{v}$}(1.87,-1);

\end{tikzpicture}
\end{center}
\caption{Choosing $\vec v$ such that it is tangent to the worldline connecting two neighbouring hyper surfaces of constant $\tau$ with identical spatial components}
\label{fig22}
\end{figure}
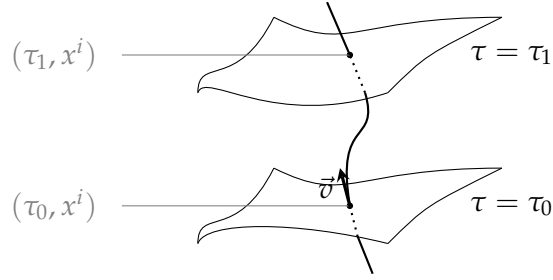

The second option is to define the time as the tangent vector to the worldline connecting two neighbouring spatial hyper surfaces with constant $\tau$ and same spatial coordinates (see figure \ref{fig22}).
In that case, normalising what just stated, we have:
\begin{align}
    e^\mu_0= \frac{1}{(-g_{ss})^{\frac{1}{2}}}\delta^\mu_s
\end{align}

Now we need to calculate and compare these dual basis to justify our choice of basis. The first option would result in the following dual components:

\begin{align}
    &e^s_0= 1-c^{-2}(a\cdot x)-\frac{1}{2}R_{0l0m}x^lx^m \nonumber \\
    &e^j_0= -c^{-1}(\omega \times x)^j +\frac{2}{3}\tensor{R}{_{0l}^j_m}x^lx^m
\end{align}

which, agrees with my decision of considering the angular velocity term to be in $e^j_0$ component. The second option, on the other hand, would result in:
\begin{align}
   &e^s_0=1+c^{-2}(a\cdot x)+\frac{1}{2}R_{0l0m}x^lx^m \nonumber\\
   &e^j_0=0
\end{align}
As it is not related to what Ito and I calculated, we will leave it.

Finally, we could say that at the level of Dirac Hamiltonian where there is no problem of having a different set of basis vectors, one could easily relate them by a Lorentz transformation. It works as follows:
\begin{align}
    e^\mu_A \Lambda^A_B = \Tilde{e}^\mu_B
\end{align}
where, $\Tilde{e}^\mu_B$ represents Ito's basis, $e^\mu_A$ is my basis and $\Lambda^A_B$ is the Lorentz transformation matrix. 
Therefore we can calculate $\Lambda^A_B$ as follows:
\begin{align}
    \Lambda^c_B= e^c_\mu \Tilde{e}^\mu_B . 
\end{align}
As the result:
\begin{align}\label{lorenz}
    &\Lambda^0_0=1 \nonumber\\
    &\Lambda^0_{\mathrm{i}}=c^{-1}(\omega \times x)_\mathrm{i} \nonumber\\
    &\Lambda^{\mathrm{i}}_0= c^{-1}( \omega \times x)^{\mathrm{i}}\nonumber\\
    &\Lambda^{\mathrm{i}}_{\mathrm{j}}=\delta^{\mathrm{i}}_{\mathrm{j}}
\end{align}
Using \eqref{lorenz} we can see that our result is transferable to Ito's result in the level of Dirac Hamiltonian,

\subsection{Comparing the Pauli Hamiltonian}

Ito calculated the Pauli Hamiltonian by using FW-transformation in terms of the physical operators. His Pauli Hamiltonian written in my convention is:
\begin{align} \label{ppp}
   H_{\text{Pauli}}&= \bigg\{ \frac{1}{2m} + \frac{1}{2mc^2}(a\cdot x)+\frac{1}{4m}R_{0l0m}x^lx^m  \bigg\}\Pi^2 \nonumber\\
   &-\frac{q}{m}\bigg\{ 1 +c^{-2}(a\cdot x)+\frac{2}{3}R_{0l0m}x^lx^m+\frac{1}{6}R_{lm}x^lx^m\bigg\} S^k B_k +\frac{1}{6m}\tensor{R}{^j_l^i_m}x^lx^m \Pi_j \Pi_i \nonumber\\
   &+\bigg\{ \frac{i}{2mc^2}a^j +\frac{1}{2mc^2}a_i S^k \tensor{\epsilon}{^{ij}_k}-\frac{i}{m}\tensor{R}{_{0l0}^j}x^l + \frac{1}{2m}\tensor{\epsilon}{^{ij}_k}S^k R_{0l0i}x^l - \frac{i}{2m}\tensor{R}{_l^j}x^l \nonumber\\
   &\quad +\frac{2c}{3}\tensor{R}{^j_{l0m}}x^lx^m + \frac{1}{4m}\tensor{\epsilon}{^{qi}_k}S^k x^l \tensor{R}{_{qil}^j}- (\omega \times x)^j \bigg\} \Pi_j \nonumber\\
   &-\frac{q}{6m}S^i B^j R_{ikjl}x^k x^l - S^i\omega_i + \frac{mc^2}{2} R_{0l0m}x^lx^m + m(a\cdot x)-\frac{1}{8m}R - \frac{1}{8m}R_{00} +\frac{ic}{6}R_{0l}x^l \nonumber\\
   &+\frac{c}{2}\tensor{\epsilon}{^{jb}_k}x^lS^kR_{bj0l}- q c A_0 +mc^2 
\end{align}
where:
\begin{align} \label{xk}
    &D_j:=i\Pi_j=\partial_j-iqA_j \quad , \text{where $\Pi$ is the canonical momentum }\\
    &\frac{\sigma^k}{2}:=S^k \quad , \text{where $S$ is spin operator}\\
    &B^i:=\frac{1}{2}\epsilon^{ijk}(\partial_jA_k-\partial_kA_j) \quad , \text{where $B$ is magnetic field}
\end{align}
On the other hand, my Pauli Hamiltonian in the same order of approximation and in terms of the mentioned physical operators would be:
\begin{align} \label{kl}
     H_{\text{Pauli}}&= \bigg\{ \frac{1}{2m} + \frac{1}{2mc^2}(a\cdot x)+\frac{1}{4m}R_{0l0m}x^lx^m  \bigg\}\Pi^2 \nonumber\\
   &-\frac{q}{m}\bigg\{ 1 +c^{-2}(a\cdot x)+\frac{1}{2}R_{0l0m}x^lx^m\bigg\} S^k B_k +\frac{1}{6m}\tensor{R}{^j_l^i_m}x^lx^m \Pi_j \Pi_i \nonumber\\
   &+\bigg\{ -\frac{i}{4mc^2}a^j +\frac{1}{2mc^2}a_i S^k \tensor{\epsilon}{^{ij}_k}+\frac{i}{12m}\tensor{R}{_{0l0}^j}x^l + \frac{1}{2m}\tensor{\epsilon}{^{ij}_k}S^k R_{0l0i}x^l + \frac{i}{3m}\tensor{R}{_l^j}x^l \nonumber\\
   &\quad +\frac{2c}{3}\tensor{R}{^j_{l0m}}x^lx^m + \frac{1}{6m}\tensor{\epsilon}{^{iq}_k}S^k x^l (\tensor{R}{_{ql}^j_i}+\tensor{R}{_{qi}^j_l})- (\omega \times x)^j \bigg\} \Pi_j \nonumber\\
   &- S^i\omega_i + \frac{mc^2}{2} R_{0l0m}x^lx^m + m(a\cdot x)+\frac{1}{8m}R + \frac{1}{4m}R_{00} +\frac{ic}{3}R_{0l}x^l \nonumber\\
   &-q c A_0+c\tensor{\epsilon}{^{jb}_k}S^k x^l (\frac{2}{3}R_{0jbl}+\frac{1}{6}R_{bj0l})
\end{align}
Due to the different choices of basis vectors and different methods of post-Newtonian approximations we cannot compare the result yet. In order to do so, one could try to make the result comparable by Lorentz transforming one Hamiltonian to the other. This might address the first difference and we will investigate this in the following section.

\subsubsection{Lorentz Transformation and relating the results}

In order to compare the result, we need to analyse how to transform Spinors which are written in different basis vectors. Using the principal fibre bundle mathematics, we know that by assuming:
\begin{align} \label{s}
\bar{e}_B^\mu=\tensor{\Lambda}{^A_B} e^\mu_A
\end{align}
Spinors transform as follows:
\begin{align} \label{q}
\bar{\psi}_{(\Tilde{e})}=\exp \left(-\frac{1}{2}\tensor{A}{^I_J} \tensor{S}{_I^J}\right)\psi_{(e)}
\end{align}
where $A$, defined by $\Lambda=\exp(A)$, is the infinitesimal Lorentz transformation and $S^{IJ}=\frac{1}{4}[\gamma^I,\gamma^J]$ is the generator of Lorentz transformation on Spinors.

As we want to perform the transformation at the level of Pauli Hamiltonian, we can use Teylor expansion in order to find what $A$ is:
\begin{align} \label{o}
\bar{e}^\mu_B &= \tensor{\Lambda}{^A_B} e^\mu_A \nonumber\\
&=\tensor{\exp(A)}{^A_B}e^\mu_A \nonumber\\
&=e^\mu_A + \tensor{A}{^A_B} e^\mu_A+\frac{\tensor{(A^2)}{^A_B}}{2}e^\mu_A
\end{align}
On the other hand:
\begin{align}
    \tensor{\Lambda}{^A_B}=e^A_\mu \bar{e}^\mu_B
\end{align}
Therefore:
\begin{align}  \label{b}
    &\tensor{\Lambda}{^0_0}=1 \nonumber\\
    &\tensor{\Lambda}{^0_{\mathrm{i}}}=c^{-1}(\omega \times x)_\mathrm{i} \nonumber\\
    &\tensor{\Lambda}{^{\mathrm{i}}_0}=c^{-1}(\omega \times x)^\mathrm{i} \nonumber\\
    &\tensor{\Lambda}{^{\mathrm{i}}_{\mathrm{j}}}=\delta^i_j
\end{align}
Now using \eqref{s} and \eqref{b} we can read off $A$ from \eqref{o}:
\begin{align}
    &\tensor{A}{^0_0}= 0 \nonumber \\
    &\tensor{A}{^i_0}=c^{-1}(\omega \times x)^i \nonumber\\
    &\tensor{A}{_i^0}=c^{-1}(\omega \times x)_i \nonumber\\
    &\tensor{A}{^i_j}=0
\end{align}
In order to complete the calculation for \eqref{q}, we need to calculate the term $\tensor{S}{_I^J}$:
\begin{subequations}
\begin{align}
    S_{IJ} &=\frac{1}{4}[\gamma_I,\gamma_J]=\frac{1}{2}\gamma_{[I} \gamma_{^J]} \\
    \tensor{S}{_I^J} &=\frac{1}{4}[\gamma_I,\gamma^J] \\
    \tensor{A}{^I_J} \tensor{S}{_I^J} &=\frac{1}{4}\bigg\{c^{-1}(\omega \times x)_i\gamma_0 \gamma^\mathrm{i}-c^{-1}(\omega \times x)_i \gamma^\mathrm{i}\gamma^0 \nonumber\\
      &\quad +c^{-1}(\omega \times x)^i\gamma_\mathrm{i}\gamma^0 -c^{-1}(\omega \times x)^\mathrm{i} \gamma^0\gamma_\mathrm{i} \bigg\} \\
    \tensor{A}{^I_J} \tensor{S}{_I^J} &= -\frac{1}{2}c^{-1}(\omega \times x)_i\gamma^0\gamma^\mathrm{i} = -\frac{1}{2}c^{-1}(\omega \times x)_i\begin{pmatrix}
  0 & \sigma^i\\ 
  \sigma^i & 0
\end{pmatrix}
\end{align}
\end{subequations}
Therefore \eqref{q} yields:
\begin{align}
    \bar{\psi}_{(\bar{e})}&=\exp\bigg\{ \frac{1}{4}c^{-1}(\omega \times x)_i\begin{pmatrix}
  0 & \sigma^i\\ 
  \sigma^i & 0   
  \end{pmatrix}\bigg \} \psi_{(e)}
\end{align}
Now we again use the Taylor series in order to apply the transformation in the level of Pauli Hamiltonian. Therefore:
 \begin{align} \label{rr}
     \bar{\psi}&=\psi + \bigg\{\frac{1}{4}c^{-1}(\omega \times x)_i\begin{pmatrix}
  0 & \sigma^i\\ 
  \sigma^i & 0   
  \end{pmatrix} \bigg\} \psi + \frac{1}{2}\bigg\{\frac{1}{4}c^{-1}(\omega \times x)_i\begin{pmatrix}
  0 & \sigma^i\\ 
  \sigma^i & 0   
  \end{pmatrix} \bigg\}^2 \psi+ O(x^3  ) \\ \nonumber
  &= \psi +  \bigg\{\frac{1}{4}c^{-1}(\omega \times x)_i\begin{pmatrix}
  0 & \sigma^i\\ 
  \sigma^i & 0   
  \end{pmatrix} \bigg\} \psi +  \bigg\{\frac{3}{32}c^{-2}(\omega \times x)^2 \mathbb{1} \bigg\} \psi
 \end{align}
Now we know that:
\begin{align}
    \psi =  \begin{pmatrix} \psi_A \\\psi_B \end{pmatrix}
\end{align}
Therefore \eqref{rr} can be simplified to the following two equations:
\begin{align}
    &\bar{\psi}_A=\psi_A + \frac{c^{-1}}{4}(\omega \times x).\sigma \psi_B + \frac{3c^{-2}}{32}(\omega \times x)^2 \psi_A \label{rrrr}\\
    &\bar{\psi}_B=\psi_B + \frac{c^{-1}}{4}(\omega \times x).\sigma \psi_A + \frac{3c^{-2}}{32}(\omega \times x)^2 \psi_B
\end{align}
Now in order to decouple the above equations, we first note that we are interested about the positive frequency solutions. Therefore we need to use the relations we found in the previous chapter for different orders of $\psi_B$. Secondly, we need to expand $\psi_B$ in the \eqref{rrrr} by its equivalent in terms of $\psi_A$ by using the equations \eqref{4.6} , \eqref{4.8} and \eqref{4.14}:
\begin{align}
    \bar{\psi}_A&=\psi_A +\frac{c^{-1}}{4}(\omega \times x).\sigma \bigg\{      \psi_B(0)+\psi_B(1)+\psi_B(2)+O(c^{-3})  \bigg\} +\frac{3c^{-2}}{32}(\omega\times x)^2 \psi_A \nonumber\\
    &=\psi_A +\frac{3c^{-2}}{32}(\omega\times x)^2 \psi_A + \frac{c^{-1}}{4}(\omega \times x).\sigma \Bigg\{ -\frac{i}{2mc}(\sigma \cdot D) \{\psi_A(0)+\psi_A(1) \}+ \nonumber\\
    &\qquad \frac{1}{12}\sigma^b R_{bl0m}x^lx^m \psi_A(0) \Bigg\} 
\end{align}
Considering the Ito's level of approximation, we can neglect the terms of the non-linear order in $\omega$ and terms of the order $c^{-3}$. Therefore we end up with:
\begin{align}
    \bar{\psi}_A= \psi_A-\frac{i}{8mc^2}(\omega \times x)_i  \sigma^i \sigma^j D_j \psi_A(0)
\end{align}
Note that in our convention $\psi_A(1)=c^{-1}\psi_A$ .
Knowing the Pauli matrices algebra \eqref{2.8}, we can simplify it further:
\begin{align} \label{5.64}
    \bar{\psi}_A&= \psi_A-\frac{i}{8mc^2}(\omega \times x)_i(\delta^{ij}+i\tensor{\epsilon}{^{ij}_k}\sigma^k) D_j \psi_A(0) \nonumber\\
    &=\psi_A -\frac{i}{8mc^2}(\omega \times x)_i D^i \psi_A(0) + \frac{1}{8mc^2}(\omega \times x )_i \tensor{\epsilon}{^{ij}_k}\sigma^k D_j \psi_A(0)
\end{align}
Now we need to replace $\psi$ in \eqref{ppp} by \eqref{5.64} and see if the resulting Pauli Hamiltonian is identical to mine as seen in \eqref{kl}.
In order to do so, we first translate the terms in \eqref{5.64} by physical operations as in \eqref{xk}:
\begin{align}
    \bar{\psi}_A&=\psi_A +\frac{1}{8mc^2}(\omega \times x)_i \Pi^i \psi_A(0) -\frac{1}{4mc^2} \tensor{\epsilon}{^{ij}_k}S^k(\omega \times x )_i \Pi_j \psi_A(0)
\end{align}
However, applying the above transformation also would not make the resulting equation comparable to mine. Therefore one would conclude that the result is not comparable and the reason is the different methods of post-Newtonian expansions which are used. 
We will briefly discuss it in the following section. 

\subsection{Different methods of post-Newtonian approximation}

It seems that it is not easy to compare the results after performing the post-Newtonian approximation. The reason is the different approaches Ito and I took to perform the post-Newtonian approximation itself. 

He is using \enquote{Foldy-Wouthuysen-like} expansion \cite{PhysRev.78.29,2020} up to the order of $\frac{1}{m}$ and I am using the formal $c$ power expansion. 
Foldy-Wouthuysen transformation (or in short FW-transformation) was first formulated by Leslie Lawrance Foldy and Siegfried Adolf Wouthuysen in 1949 to study the non-relativistic limit of the Dirac equation in the flat background \cite{PhysRev.78.29,PhysRev.87.688,1951PThPh...6..267T}. Ever since physicists mostly use it in order to talk about the non-relativistic limit. It is a unitary transformation of the orthonormal basis in which both the Hamiltonian and the state are represented. The eigenvalues do not change under such a unitary transformation, this means the physics does not change under such a unitary basis transformation. Therefore such a unitary transformation can always be applied. In particular a unitary basis transformation may be picked which will put the Hamiltonian in a more pleasant form, at the expense of a change in the state function, which then represents something else \cite{fetter2012quantum}. 
It, however, seems to be problematic. The first reason is that it is not clear whether FW-transformation is well-defined in the context of curved spacetime. Secondly, as  scientist want to interpret the terms appearing in the Hamiltonian to set up an experiment and observe them, they will face a problem. Because the unitary transformation makes the Hamiltonian to be dependant on the taste of the person who does the calculation. As stated earlier, the unitary basis transformation may be picked which will put the Hamiltonian in a more pleasant form. Therefore, one could not decide whether the terms appearing in the Hamiltonian correspond to a realistic effect which is detectable in the laboratory.

On the other hand, the method which I have shown in this thesis, is pretty much independent of the taste of the person who performs the calculation. It is a straightforward, mathematically-consistent and logical process which provides us not only with the non-relativistic limit, but also with the post-Newtonian corrections.  
Physical operators such as Spin and canonical momentum operators are the corresponding lab measurement tools in real world experiments. In order to define them properly one needs to be really careful. 
It is the other objection to Ito's choice of basis vectors and his method of non-relativistic expansion. Because, it is not clear at all why he thinks that the spin operator would still be defined as $S=\sigma/2$ after performing the FW-transformation. 
In other words, the axis of the rotation would have been differed, if one would choose another set of basis vectors. That means the interpretation of the terms and coupling in the Hamiltonian is impossible.

\chapter{Conclusion} \label{chap:concl}

We have achieved the description of spin-half particle under gravity also known as the Pauli equation with corrections. We believe our approach to be the only systematic and complete method for the inclusion of gravitational effects, i.e., the inclusion of \emph{all} the effects without repetition, other than in existing literature \cite{Zhen_Hua_2007,DAI20163601,PhysRevD.76.064016,Arminjon_2006}. 
Our calculation consisted of two steps: first, we expanded the Dirac Hamiltonian in the FNC in curved spacetime \eqref{eq:Dirac_FNC}, which implemented certain approximations based on the assumption of weak gravity and weak inertial effects. Second, we studied the resulting Hamiltonian in the post-Newtonian regime \eqref{eq4.25}, which implemented the assumption of slow velocities. We note that these two steps of our calculation are logically independent because they rely on different sets of self-standing assumptions.  Our work expands on the existing literature \cite{Ito:2020xvp,perche21} by systematic inclusion of neglected terms.

The mentioned resulting Hamiltonians are significant findings, because they can be applied in different research areas ranging from theoretical quantum gravity to experimental quantum optics and atom interferometry. Additionally, my results can be applied to all the following ongoing research:
\begin{itemize}
    \item Electron $g$-factor measurements \cite{Ito:2020xvp,PhysRevLett.100.120801,PhysRevLett.97.030801}
    \item Localisation of fermions in AdS backgrounds \cite{perche21}
    \item Unruh effect studies \cite{PhysRevD.14.870,PhysRevD.7.2850,Davies_1975,PhysRevLett.91.243004,2021,2019:nature,2018:martinez} 
    \item Atoms in weak gravitational fields \cite{doi:10.1098/rspa.1928.0023,PhysRevLett.118.183602,PhysRevLett.120.183604,PhysRevD.101.125018,PhysRevLett.114.013001}
\end{itemize}

\enlargethispage{\baselineskip}

A possible further research direction would be to design novel experiments for observing the coupling effects derived from my resulting Hamiltonians.
One could also make the situation more realistic by dropping the assumption of time independence in curvature, acceleration and angular velocity.
Another potentially interesting topic for further research is the relationship between the description of a Dirac field in curved spacetime and the description of classical spinning bodies by the Mathisson--Papapetrou--Dixon equations: this has been somewhat explored in the fully relativistic case \cite{Audretsch_1981}, but we believe that exploring it via post-Newtonian expansions may lead to further understanding.

\appendix
\chapter{Calculation of inverse metric} \label{appendix A}

In order to calculate the inverse metric, as the off-diagonal components $g_{si}$ does not vanish, we should first separate the spatial components of the metric and rename it as follows:
 \begin{align}
     g_{ij}=:h_{ij}
 \end{align}
 for which the inverse , namely $h^{ij}$ , can be calculated by using the geometric series.
Then define: 
\begin{align}
    \beta^i:=h^{ij}g_{sj} \quad \text{and} \quad \alpha^2:=h_{ij}\beta^i\beta^j-g_{ss}
\end{align}
We note that:
\begin{align}
    h_{ij}\beta^i\beta^j=h^{ij}g_{si}g_{sj}
\end{align}
Now it can be proven that:
\begin{align}
    &g^{ss}=-\alpha^{-2} \\
    &g^{si}=\alpha^{-2}\beta^i \\
    &g^{ij}=h^{ij}-\alpha^{-2}\beta^i\beta^j
\end{align}
Therefore we need to calculate $h^{ij}$ , $\beta^i$ and $\alpha^2$ to be able to calculate the inverse metric.
Given the recipe, we can start by calculating the $h^{ij}$: 
\begin{align}
    h_{ij}=\delta_{ij}-\frac{1}{3}R_{iljm}x^lx^m 
\end{align}
Using the geometric series:
\begin{align} \label{a.11}
    \frac{1}{1-x}=1+x+x^2+...
\end{align}
Therefore:
\begin{align}
        h^{ij}=\delta^{ij}+\frac{1}{3}\tensor{R}{^i_l^j_m}x^lx^m
\end{align}
Now for $\beta^i$ we have:
\begin{align}
    \beta^i&=h^{ij}g_{sj} \nonumber\\
    &=\bigg\{\delta^{ij}+\frac{1}{3}\tensor{R}{^i_l^j_m}x^lx^m\bigg\} \bigg\{c^{-1}(\omega\times x)_j-\frac{2}{3}R_{0ljm}x^lx^m\bigg\} \nonumber\\ 
    &=c^{-1}(\omega\times x)^i-\frac{2}{3}\tensor{R}{_{0l}^i_m}x^lx^m
    \end{align}
and, finally for the $\alpha^2$ we have:
\begin{align}
    \alpha^2&=h^{ij}g_{si} g_{sj}-g_{ss} \nonumber\\
    &=\big\{\delta^{ij}+\tfrac{1}{3} \tensor{R}{^i_l^j_m}x^lx^m\big\}\big\{c^{-1}(\omega\times x)_i - \tfrac{2}{3}R_{0lim}x^lx^m\big\}\big\{c^{-1}(\omega\times x)_j-\tfrac{2}{3}R_{0ljm}x^lx^m\big\} - g_{ss}\nonumber\\
    &=\delta^{ij}c^{-2}(\omega \times x)_i(\omega \times x)_j -g_{ss}\nonumber\\
    &=c^{-2}(\omega\times x)^2-g_{ss} \nonumber\\
    &=1+2c^{-2}(a\cdot x)+c^{-4}(a\cdot x)^2 +R_{0l0m}x^lx^m
\end{align}
Now in order to calculate $\alpha^{-2}$ we use the geometric series:
\begin{align}
    g^{ss}&=-\alpha^{-2} \nonumber\\
    &=-(1+2c^{-2}(a\cdot x)+c^{-4}(a\cdot x)^2+R_{0l0m}x^lx^m)^{-1} \nonumber\\
    &=-\bigg\{1-2c^{-2}(a\cdot x)-c^{-4}(a\cdot x)^2-R_{0l0m}x^lx^m\nonumber\\ &\qquad +\{\ 2c^{-2}(a\cdot x)+c^{-4}(a\cdot x)^2 +  R_{0l0m}x^lx^m\}^2 + O(x^3)\bigg\} \nonumber\\
    &=-(1-2c^{-2}(a\cdot x)-c^{-4}(a\cdot x)^2-R_{0l0m}x^lx^m+4c^{-4}(a\cdot x)^2) \nonumber\\
    &=-(1-2c^{-2}(a\cdot x)+3c^{-4}(a\cdot x)^2-R_{0l0m}x^lx^m)
\end{align}

\begin{align}
    g^{si}&=\alpha^{-2}\beta^i\nonumber\\
    &=\bigg\{1-2c^{-2}(a\cdot x)+3c^{-4}(a\cdot x)^2-R_{0l0m}x^lx^m\bigg\}\bigg\{c^{-1}(\omega\times x)^i-\frac{2}{3}\tensor{R}{_{0l}^i_m}x^lx^m\bigg\}\nonumber\\
    &=c^{-1}(\omega\times x)^i -\frac{2}{3}\tensor{R}{_{0l}^i_m}x^lx^m -2c^{-3}(a\cdot x)(\omega \times x)^i
\end{align}
and, finally similar to previous calculation:
\begin{align}
    g^{ij}=\delta^{ij}+\frac{1}{3}\tensor{R}{^i_l^j_m}x^lx^m-c^{-2}(\omega \times x)^i (\omega\times x)^j
\end{align}
At the end, we can calculate the inverse of the metric $(g^{ss})^{-1}$ by using the geometric series \eqref{a.11} up to the order of $x^3$.
\begin{align}
    (g^{ss})^{-1}&=-\{1-2c^{-2}(a\cdot x)+3c^{-4}(a\cdot x)^2-R_{0l0m}x^lx^m\}^{-1}\nonumber\\
    &=-\bigg\{1+\{ 2c^{-2}(a\cdot x)-3c^{-4}(a\cdot x)^2+R_{0l0m}x^lx^m\}\nonumber \\
    &\qquad+\{ 2c^{-2}(a\cdot x)-3c^{-4}(a\cdot x)^2+R_{0l0m}x^lx^m\}^{2}\bigg\}\nonumber\\
    &=-\bigg\{1+\{ 2c^{-2}(a\cdot x)-3c^{-4}(a\cdot x)^2+R_{0l0m}x^lx^m\}+\{ 4c^{-4}(a\cdot x)^2\}\bigg\}\nonumber\\
    &=-1-2c^{-2}(a\cdot x)-c^{-4}(a\cdot x)^2-R_{0l0m}x^lx^m
\end{align}
It is needed in section \ref{oks} in order to calculate the Dirac Hamiltonian which is expanded in the FNC.

\chapter{Calculation of Christoffel symbols} \label{appendix B}

In order to calculate the Christoffel symbols, we will use the definition:
\begin{align}
\tensor{\Gamma}{^\mu_{\nu\rho}}=\frac{1}{2}g^{\mu \kappa}(g_{\kappa\nu,\rho}+g_{\kappa\rho,\nu}-g_{\nu\rho,\kappa})
\end{align}
In this case, we need to calculate the following 6 components:
$\tensor{\Gamma}{^s_{ij}}$ , $\tensor{\Gamma}{^s_{si}}$ , $\tensor{\Gamma}{^i_{ss}}$ , $\tensor{\Gamma}{^s_{ss}}$ ,  $\tensor{\Gamma}{^i_{jk}}$ and $\tensor{\Gamma}{^i_{sj}}$.
\begin{align}
 \tensor{\Gamma}{^s_{ij}}&=\frac{1}{2}g^{ss}(g_{si,j}+g_{sj,i}-g_{ij,s}) +\frac{1}{2}g^{sn}(g_{ni,j}+g_{nj,i}-g_{ij,n})   \nonumber\\
 &=\frac{1}{2}g^{ss}\bigg\{c^{-1}(\omega \times x)_{i,j}-\frac{2}{3}R_{0lim}(x^l x^m)_{,j}+c^{-1}(\omega \times x)_{j,i}-\frac{2}{3}R_{0ljm}(x^l x^m)_{,i}\bigg\} \nonumber\\
 &\quad +\frac{1}{2}g^{sn}\bigg\{ (\delta_{ni}-\frac{1}{3}R_{nlim}x^lx^m)_{,j}+(\delta_{nj}-\frac{1}{3}R_{nljm}x^lx^m)_{,i}-(\delta_{ij}-\frac{1}{3}R_{iljm}x^lx^m)_{,n} \bigg\}
\end{align}
Note that throughout this thesis, the time derivative of $R$, $a$ and $\omega$ are ignored. Therefore terms such as $g_{ij,s}$ are vanishing.
Now in order to further simplify the expression we use the definition of the vector product:
\begin{align}
     \tensor{\Gamma}{^s_{ij}}&=
 \frac{1}{2}g^{ss}\bigg\{c^{-1} \tensor{\epsilon}{_{iqk}}\omega^q x^k_{\;,j}-\frac{2}{3}R_{0lim}(x^l_{\;,j} x^m+x^l x^m_{\;,j}) \nonumber\\
 &\qquad+ c^{-1} \tensor{\epsilon}{_{jqk}}\omega^q x^k_{\;,i}-\frac{2}{3}R_{0ljm}(x^l_{\;,i} x^m+x^l x^m_{\;,i})\bigg\} \nonumber\\
 &\quad +\frac{1}{6}g^{sn} \bigg\{{-R_{nlim}}(x^l_{\;,j} x^m+x^l x^m_{\;,j}) - R_{nljm}(x^l_{\;,i} x^m+x^l x^m_{\;,i}) + R_{iljm}(x^l_{\;,n} x^m+x^l x^m_{\;,n})) \bigg\}
\end{align}
Now we know that $x^m_{\;,n}={\delta}{^m_n}$ . Moreover, using the fact that $\tensor{\epsilon}{_{iqk}}$ is totally anti-symmetric we can simplify the expression as follows:
\begin{align}
     \tensor{\Gamma}{^s_{ij}}&=
 -\frac{1}{3}g^{ss}\bigg\{(R_{0kil}+R_{0lik}){\delta}{^k_j}x^l+(R_{0kjl}+R_{0ljk}){\delta}{^k_i}x^l\bigg\} \nonumber\\
 &\quad -\frac{1}{6}g^{sn}\bigg\{ (R_{nkil}+R_{nlik}){\delta}{^k_j}x^l+(R_{nkjl}+R_{nljk})\tensor{\delta}{^k_i}x^l-(R_{ikjl}+R_{iljk}){\delta}{^k_n}x^l \bigg\} 
\end{align}
Now we can use the relations of Riemann tensor. Therefore: 
\begin{align}
     \tensor{\Gamma}{^s_{ij}}&=
 -\frac{1}{3}g^{ss}\bigg\{ (R_{0jil}+R_{0ijl})x^l\bigg\}
-\frac{1}{3}g^{sn}\bigg\{ (R_{njil}+R_{nijl})x^l \bigg\} 
\end{align}
Opening $g^{ss}$ and $g^{sn}$ and multiplying the terms up to and including the order of $x^2$ will give us the final result:
\begin{align}
     \tensor{\Gamma}{^s_{ij}}=\frac{1}{3}x^l\bigg \{ (R_{0jil}+R_{0ijl})-2c^{-2}(a\cdot x)(R_{0jil}+R_{0ijl})-c^{-1}(\omega\times x)^n (R_{njil}+R_{nijl})\bigg\}
\end{align}
The rest can be calculated the same way. 

\chapter{Calculation of the connection one-forms} \label{appendix C}

Connection one-forms can be calculated by:
\begin{align} \label{goy}
    \tensor{w}{_\mu ^I_J}= \tensor{\Gamma}{^\alpha_{\mu \nu}}e^I_\alpha e^\nu_J -e^\nu_J \partial_\mu e^I_\nu
\end{align}
As we already have all the required terms for the above equation, we can calculate the connection one-forms components to components. For instance: 
\begin{align}\label{zozol}
    \tensor{w}{_s^0_0}=\tensor{\Gamma}{^s_{ss}}e^0_s e^s_0+\tensor{\Gamma}{^i_{ss}}e^0_i e^s_0 +\tensor{\Gamma}{^s_{sj}}e^0_s e^j_0+\tensor{\Gamma}{^i_{sj}}e^0_i e^j_0-e^s_0\partial_s e^0_s -e^j_0 \partial_s e^0_j
\end{align}
In order to calculate \eqref{zozol}, we study it term by term.
The first term:
\begin{align}
\tensor{\Gamma}{^s_{ss}} e^0_s e^s_0 = \tensor{\Gamma}{^s_{ss}} \{1+c^{-2}(a\cdot x) - \tfrac{1}{2}\tensor{R}{^0_{l0m}}x^lx^m\} \{1-c^{-2}(a\cdot x)+c^{-4}(a\cdot x)^2 - \tfrac{1}{2}R_{0l0m}x^lx^m\}
\end{align}
So far, we just replaced the $e^0_s$ and $e^s_0$ by their values in \eqref{slm} and \eqref{salam}. Note that we need to keep the order up and including $x^2$. Therefore we can simplify the above expression by multiplying the brackets:
\begin{align}
\tensor{\Gamma}{^s_{ss}}e^0_s e^s_0=\tensor{\Gamma}{^s_{ss}}\bigg\{1-c^{-2}(a\cdot x)+c^{-4}(a\cdot x)^2-\frac{1}{2}R_{0l0m}x^lx^m+ \nonumber \\
\qquad c^{-2}(a\cdot x)-c^{-4}(a\cdot x)^2 -\frac{1}{2}\tensor{R}{^0_{l0m}}x^lx^m\bigg\}
\end{align}
Now we know that lowering the time-indices, due to our convention and the fact that the frame is Minkowski metric, will give us a minus sign. I.e:
\begin{align*}
  \tensor{R}{^0_{l0m}}=-R_{0l0m}  
\end{align*}
Therefore:
\begin{align}
\tensor{\Gamma}{^s_{ss}}e^0_s e^s_0=\tensor{\Gamma}{^s_{ss}}
\end{align}
For the second term, we will use the same approach:
\begin{align}
\tensor{\Gamma}{^i_{ss}}e^0_i e^s_0 &=\tensor{\Gamma}{^i_{ss}}\bigg\{-\frac{1}{6}\tensor{R}{^0_{lim}}x^lx^m   \bigg\}\bigg\{ 1-c^{-2}(a\cdot x)+c^{-4}(a\cdot x)^2-\frac{1}{2}R_{0l0m}x^lx^m  \bigg\} \nonumber\\
&=\tensor{\Gamma}{^i_{ss}}(-\frac{1}{6}\tensor{R}{^0_{lim}}x^lx^m) \nonumber\\
&=\frac{1}{6}c^{-2}a^iR_{0lim}x^lx^m
\end{align}
where, in the last line we replaced $\tensor{\Gamma}{^i_{ss}}$ by its value in \eqref{laklak}.
The third term:
\begin{align}
    \tensor{\Gamma}{^s_{sj}}e^0_s e^j_0&= \tensor{\Gamma}{^s_{sj}} \bigg\{1+c^{-2}(a\cdot x)+\frac{1}{2}R_{0l0m}x^lx^m  \bigg\} \nonumber \\ & \qquad \bigg\{-c^{-1}(\omega \times x)^j +c^{-3}(a\cdot x)(\omega \times x)^j +\frac{1}{2}\tensor{R}{_{0l}^j_m}x^lx^m \bigg\} \nonumber\\
    &=\tensor{\Gamma}{^s_{sj}}(-c^{-1}(\omega\times x)^j+\frac{1}{2}\tensor{R}{_{0l}^j_m}x^lx^m) \nonumber\\
    &=-c^{-3}a.(\omega\times x)+c^{-5}a.(\omega\times x)(a\cdot x)-c^{-1}(\omega \times x)^jR_{0j0l}x^l+\frac{1}{2}c^{-2}a_j\tensor{R}{_{0l}^j_m}x^lx^m
\end{align}
The forth term:
\begin{align}
    \tensor{\Gamma}{^i_{sj}}e^0_i e^j_0&= \tensor{\Gamma}{^i_{sj}}\bigg\{ \frac{1}{6}R_{0lim}x^lx^m \bigg\} \bigg\{ -c^{-1}(\omega \times x)^j +c^{-3}(a\cdot x)(\omega \times x)^j +\frac{1}{2}\tensor{R}{_{0l}^j_m}x^lx^m \bigg\} \nonumber\\
    &=0
\end{align}
Again, in the last line we ask our approximation to be up to and including the order of $x^2$ and that is why we end up to zero. 
Finally, the last two terms are also zero because of our assumption that the time variation of our variables are negligible. 
All in all, we will insert each terms back into \eqref{zozol}:
\begin{align}\label{zozo}
    \tensor{w}{_s^0_0}&=\tensor{\Gamma}{^s_{ss}}e^0_s e^s_0+\tensor{\Gamma}{^i_{ss}}e^0_i e^s_0 +\tensor{\Gamma}{^s_{sj}}e^0_s e^j_0+\tensor{\Gamma}{^i_{sj}}e^0_i e^j_0-e^s_0\partial_s e^0_s -e^j_0 \partial_s e^0_j \nonumber\\
    &=\tensor{\Gamma}{^s_{ss}}+\frac{1}{6}c^{-2}a^iR_{0lim}x^lx^m-c^{-3}a.(\omega\times x)+c^{-5}a.(\omega\times x)(a\cdot x)\nonumber\\
    &\quad-c^{-1}(\omega \times x)^jR_{0j0l}x^l+\frac{1}{2}c^{-2}a_j\tensor{R}{_{0l}^j_m}x^lx^m
\end{align}
Using \eqref{koso} to replace $\tensor{\Gamma}{^s_{ss}}$, we end up:
\begin{align}
    \tensor{w}{_s^0_0}&= c^{-3}(a.(\omega \times x))- c^{-5}(a.(\omega \times x))(a\cdot x)+c^{-1}R_{0n0l} x^l (\omega \times x)^n\nonumber\\
   & \quad-\frac{2}{3}c^{-2}a_n\tensor{R}{_{0l}^n_m}x^lx^m+\frac{1}{6}c^{-2}a^iR_{0lim}x^lx^m-c^{-3}a.(\omega\times x)\nonumber\\
   & \quad+c^{-5}a.(\omega\times x)(a\cdot x)-c^{-1}(\omega \times x)^jR_{0j0l}x^l+\frac{1}{2}c^{-2}a_j\tensor{R}{_{0l}^j_m}x^lx^m
\end{align}
Renaming the indices, we will see that they are all cancelling each other out. 
Therefore:
\begin{align}
    \tensor{w}{_s^0_0}=0
\end{align}
The others are also calculated with the same approach. As the calculation is rather straightforward and tedious I will not write them down.

\chapter{Calculation of spin connection} \label{appendix D}

The spin connection is to be calculated by:
\begin{align}
    \Gamma_\mu =-\frac{1}{2} w_{\mu IJ}S^{IJ}
\end{align}
where, $S^{IJ}$ is given by \eqref{plpl}. Therefore:
\begin{align}
    \Gamma_\mu =-\frac{1}{8} w_{\mu IJ}[\gamma^I , \gamma^J]
\end{align}
where:
\begin{align*}
    [\gamma^I , \gamma^J]=\gamma^I\gamma^J - \gamma^J \gamma^I
\end{align*}
is the commutation relation.
Now we break the spin connection into its temporal and spatial parts. We have the following:
\begin{align}
\Gamma_s &=-\frac{1}{8} w_{s IJ}[\gamma^I , \gamma^J]\nonumber\\
&=-\frac{1}{8}\bigg\{w_{s00}[\gamma^0,\gamma^0]+w_{s0j}[\gamma^0, \gamma^j] +w_{si0}[\gamma^i, \gamma^0]+w_{sij}[\gamma^i, \gamma^j]\bigg\} \nonumber\\
&=-\frac{1}{8}\bigg\{w_{s0i}[\gamma^0, \gamma^i] -w_{s0i}[\gamma^i, \gamma^0]+w_{sij}[\gamma^i, \gamma^j]\bigg\}
\end{align}
where, in the last line we Note that we renamed the index in $w_{s0j}[\gamma^0, \gamma^j]$ and also we used the anti-symmetric property of $w$. I.e.\ $w_{s0i}=-w_{si0}$.
Therefore we can simplify it:
\begin{align}
\Gamma_s &=-\frac{1}{8}\bigg\{2 w_{s0i}[\gamma^0, \gamma^i] +w_{sij}[\gamma^i, \gamma^j]\bigg\} \nonumber\\
&=-\frac{1}{8}\bigg\{4 w_{s0i}\gamma^0 \gamma^i +2w_{sij}\gamma^i \gamma^j\bigg\} \nonumber\\
&=-\frac{1}{2}w_{s0i}\gamma^0\gamma^i-\frac{1}{4}w_{sij}\gamma^i \gamma^j
\end{align} 
where, in the second line we again use the anti-symmetry property of both commutators and connection one-forms in order to drop the commutation brackets. 
Now we just need to replace each term by using our previous calculations in \eqref{dodo} and \eqref{dododo}. Therefore we end up with:
\begin{align}
    &\Gamma_s= \frac{1}{2} \gamma^0 \gamma^{\mathrm{i}} \bigg\{ c^{-2}a_i +R_{0i0l}x^l -\frac{c^{-2}}{2}a_i R_{0l0m}x^lx^m -c^{-2}(a\cdot x)R_{0i0l}x^l -c^{-1}(\omega \times x)^n x^l R_{0lni} \nonumber \\
    &\qquad +\frac{c^{-2}}{6}a_j \tensor{R}{_{il}^j_m}x^lx^m -\frac{c^{-1}}{2} \tensor{\epsilon}{^n_{pi}}\omega^p R_{0lnm}x^lx^m \bigg\} \nonumber\\
    &\quad-\frac{1}{4}\gamma^{\mathrm{i}}\gamma^{\mathrm{j}} \bigg\{ \frac{c^{-1}}{6}\omega^p x^lx^m (\epsilon^{n}_{\;\;pj}R_{ilnm}+\epsilon^{\;\;\;n}_{ip}R_{jlnm}) +c^{-1}\epsilon_{ipj}\omega^p -R_{0lij}x^l \nonumber\\
    &\qquad +\frac{c^{-2}}{6}x^lx^m(a_j R_{il0m}-a_iR_{jl0m}) \bigg\}
\end{align}
The same approach will be adopted to calculate the spatial part.

\chapter{Some tricks} \label{appendix E}

In the equation \eqref{4.21} we can do the following simplification for the term
\begin{align*}
    \frac{1}{4m^2c^4} \{-ic\sigma^j D_j \} \{-iD_\tau\} \{-ic\sigma^j D_j \} \Tilde{\psi}_A(0) 
\end{align*}  
as follows:
\begin{align}\label{okokok}
&=(\frac{i} {4m^2c^2}) \sigma^b \sigma^j D_b D_\tau D_ j \Tilde{\psi}_A(0) \nonumber\\
&= \frac{i}{4m^2 c^2} \sigma^b \sigma^j D_b [D_\tau,D_j]\Tilde{\psi}_A(0) +  \frac{1}{4m^2 c^2} (\sigma \cdot D)^2 H(0) \Tilde{\psi}_A(0) +\frac{q}{4m^2c^2} (\sigma \cdot D)^2 (c A_0 \Tilde{\psi}_A(0))
\end{align}
in which, the first term is related to electromagnetic fields. The second term is however calculated by the fact that $iD_\tau\Tilde{\psi}_A(0)= H(0)\Tilde{\psi}_A(0) +q c A_0 \Tilde{\psi}_A(0)$. The first term can be further simplified as follows:
\begin{align}
    &\hspace{-2ex}\frac{i}{4m^2c^2}\sigma^b \sigma^j D_b[ D_\tau, D_j] \nonumber\\
    &=\frac{i}{4m^2c^2}\sigma^b \sigma^j D_b\Bigg\{(\partial_\tau-iq c A_0)(\partial_j-iq A_j)-(\partial_j-iq A_j)(\partial_\tau-iq c A_0) \Bigg\} \nonumber\\
    &=\frac{i}{4m^2c^2}\sigma^b \sigma^j D_b \bigg\{iq(c \partial_j A_0 - \partial_\tau A_j) \bigg\}\nonumber\\
    &= - \frac{q}{4m^2c^2}\sigma^b \sigma^j D_b E_j 
\end{align}
Note that in our convention, $D_bE_j$ is acting on the $\psi$ and derivative needs to be expanded on both $E_j$ and $\psi$ in the resulting Hamiltonian.

We can also simplify the second term in \eqref{okokok} as follows:
\begin{align}
    \frac{1}{4m^2c^2}(\sigma \cdot D)^2 H(0) &= \frac{1}{4m^2c^2}(\sigma \cdot D)^2\Bigg\{ -\frac{1}{2m} (\sigma \cdot D)^2 +m(a\cdot x) + \frac{mc^2}{2}R_{0l0m}x^l x^m \nonumber\\
    &\qquad \qquad \qquad \qquad +i (\omega \times x)^i D_i - \frac{1}{2} \sigma_p \omega^p -q c A_0\Bigg\} \nonumber\\
    &=-\frac{1}{8m^3c^2}(\sigma \cdot D)^4 + \frac{1}{4mc^2}(\sigma \cdot D)^2(a\cdot x)\nonumber\\
    &\quad+\frac{1}{8m}R_{0l0m}(\sigma \cdot D)^2(x^lx^m)+\frac{i}{4m^2c^2}(\sigma \cdot D)^2(\omega \times x)^i D_i \nonumber\\
    &\quad -\frac{1}{8m^2c^2}\omega^p(\sigma \cdot D)^2\sigma_p -\frac{q}{4m^2c^2}(\sigma \cdot D)^2 c A_0
\end{align}
where, we used \eqref{hami0} to replace $H(0)$.
Note that the last term will cancel the last term in \eqref{okokok}. Now we need to apply the derivative to each term and simplify:
\begin{align}
    \frac{(\sigma \cdot D)^2}{4m^2c^2} H(0)&=-\frac{1}{8m^3c^2}(\sigma \cdot D)^4 + \frac{1}{4mc^2}(\sigma \cdot D)^2(a\cdot x)+\frac{1}{8m}R_{0l0m}(\sigma \cdot D)^2(x^lx^m) \nonumber\\
    &\quad + \frac{i}{4m^2c^2}(\sigma \cdot D)^2(\omega \times x)^i D_i -\frac{1}{8m^2c^2}\omega^p(\sigma \cdot D)^2\sigma_p -\frac{q}{4m^2c^2}(\sigma \cdot D)^2 c A_0 \nonumber\\
    &=-\frac{1}{8m^3c^2}(\sigma \cdot D)^4 + \frac{1}{4mc^2}(a\cdot x)(\sigma \cdot D)^2+\frac{1}{4mc^2}a_i \{\sigma^i,\sigma^j\}D_j \nonumber\\
    &\quad + \frac{1}{8m}R_{0l0m}x^lx^m(\sigma \cdot D)^2+\frac{1}{4m}R_{0i0m}x^m\{\sigma^i ,\sigma^j\}D_j+\frac{1}{8m}R_{0i0j}\{\sigma^i , \sigma^j \} \nonumber\\
    &\quad + \frac{i}{4m^2c^2}\tensor{\epsilon}{_{mn}^i}\omega^m \{\sigma^p ,\sigma^n \}D_pD_i + \frac{i}{4m^2c^2}(\omega \times x)^i (\sigma \cdot D)^2D_i \nonumber\\
    &\quad - \frac{1}{8m^2c^2}\sigma^i \sigma^j \sigma_p \omega^p D_i D_j -\frac{q}{4m^2c^2}(\sigma \cdot D)^2 c A_0
\end{align}
Note that again in our convention $(\sigma \cdot D)^2=\sigma^iD_i \sigma^j D_j$ is just for the sake of simplicity.\footnote{It can be also simplified as $(\sigma \cdot D)^2 = \sigma^i \sigma^j D_i D_j = (\delta^{ij}+i\tensor{\epsilon}{^{ij}_k}\sigma^k)D_i D_j = D^2 + i\sigma\cdot(D \times D) = D^2 + q(\sigma \cdot B)$,
where for the second equality we used the Pauli matrices algebra.}
In addition, $\{$ and $\}$ denote anti-commutators.
Finally, as we know $\{\sigma^i, \sigma^j\}=2\delta^{ij}$ we can simplify the above expression as:
\begin{align}
       \frac{1}{4m^2c^2}(\sigma \cdot D)^2 H(0)&= -\frac{1}{8m^3c^2}(\sigma \cdot D)^4 + \frac{1}{4mc^2}(a\cdot x)(\sigma \cdot D)^2+\frac{1}{2mc^2}a^j D_j+\nonumber\\
    &\quad \frac{1}{8m}R_{0l0m}x^lx^m(\sigma \cdot D)^2+\frac{1}{2m}\tensor{R}{_0^j_{0m}}x^mD_j+\frac{1}{4m}R_{00}+\nonumber\\
    &\quad \frac{i}{2m^2c^2}\tensor{\epsilon}{_{mn}^i}\omega^m D^nD_i + \frac{i}{4m^2c^2}(\omega \times x)^i (\sigma \cdot D)^2D_i-\nonumber\\
    &\quad \frac{1}{8m^2c^2}\sigma^i \sigma^j \sigma_p \omega^p D_i D_j -\frac{q}{4m^2c^2}(\sigma \cdot D)^2 c A_0
\end{align}
This can be replaced as the second term in \eqref{okokok} and can be inserted back to \eqref{4.21} in order to calculate $H(2)$ in \eqref{4.23}.

As another calculation tricks, in the equation \eqref{eq4.25} considering the term:
\begin{align*}
\frac{i}{4m^2c^2} \sigma^i \sigma^j (\omega \times x)^k D_i D_j D_k
\end{align*}
we can further simplify it as follows:
\begin{align}
\frac{i}{4m^2c^2} \sigma^i \sigma^j (\omega \times x)^k D_i D_j D_k&=\frac{i}{4m^2c^2}(\delta^{ij}+i\tensor{\epsilon}{^{ij}_m}\sigma^m)(\omega\times x)^k D_iD_jD_k \nonumber\\
&=\frac{i}{4m^2c^2}(\omega\times x)^k\bigg\{ D^2+i \sigma.(D\times D)\bigg\}D_k \nonumber\\
&=\frac{i}{4m^2c^2}(\omega\times x)^k\bigg\{ D^2-i \sigma.iqB\bigg\}D_k \nonumber\\
&=\frac{i}{4m^2c^2}(\omega\times x)^k\bigg\{ D^2+q \sigma \cdot B\bigg\}D_k 
\end{align}
where, $B$ is the magnetic field.
Same approach was employed in order to calculate the terms related to $E$ in \eqref{eq4.25}.

\printbibliography

\end{document}
%